\documentclass[12pt]{article}
\usepackage{graphicx}

\newcommand{\be}{\begin{equation}}
\newcommand{\ee}{\end{equation}}
\newcommand{\bea}{\begin{eqnarray}}
\newcommand{\eea}{\end{eqnarray}}
\newcommand{\ba}{\begin{array}}
\newcommand{\ea}{\end{array}}

\newcommand{\slashs}[1]{\not{\!#1}}

\newcommand{\norsl}{\normalsize\sl}
\newcommand{\norsc}{\normalsize\sc}

\newcommand{\NB}[1]{\langle #1 \rangle}
\newcommand{\PB}[1]{[#1]}

\begin{document}

%\begin{titlepage}

\title{ \bf
Five-parton amplitudes with 
two-quark and two-photon
at Next-to-leading Order
}

\author{
\norsc  Yoshiaki YASUI\\
\norsl
KEK \\
Oho 1-1, Tsukuba,Ibaraki  305-0801, Japan\\ }

\date{
 }

\maketitle

\abstract{
We discuss one-loop five-parton amplitudes  
with two-quark two-photon or three-photon external legs. 
The amplitudes are required to evaluate the NLO 
corrections for the $\gamma\gamma jet$ production process
at hadron colliders. 
The results have been already discussed by 
several groups in terms of QCD amplitudes. 
Here, we present more straightforward version 
of one-loop calculations without using the QCD amplitudes. 
}

\begin{picture}(0,0)(-350,-250)
\put(0,0){KEK-TH-811}
\end{picture}

%\baselineskip 24pt

%\newpage
%\input chap1.tex
\section{Introduction}

The search for the Higgs boson will be one of the 
most important issue  of the LHC experiment. 
However, if the mass of the Higgs boson  is light 
($M_H \le 140$ GeV), 
the Higgs boson search at the LHC is not so easy. 
In this range, the rare decay mode into two photons 
($H\rightarrow\gamma\gamma$) is expected to 
give rather clear signals\cite{TDR}. 
There have been many calculations for this process 
including the radiative corrections \cite{SPIRA,HB}. 
They found that the next-to-leading order (NLO) corrections to 
the Higgs production are very large\cite{SPIRA}. 
The next-to-next-to-leading order (NNLO) corrections 
haven been evaluated in the large top mass limit\cite{HB}. 
The NNLO corrections show a good convergence 
in the perturbative expansion. 
Now, the uncertainty of the higher order corrections 
for the Higgs production signal is less serious. 
On the other hand, the QCD background still 
has large uncertainties. 
The NLO corrections of the QCD background 
($q\bar{q}\rightarrow \gamma\gamma$) are also very large. 
To make matters worse, NNLO corrections of 
gluon-gluon initiated processes, 
such as the Box type correction 
$gg\rightarrow\gamma\gamma$, may very large 
due to the large gluon distribution\cite{FLO,BAL}. 
Actually, the size of a NNLO correction $gg\rightarrow \gamma\gamma$ 
 comparable to the LO contribution $q\bar{q}\rightarrow \gamma\gamma$. 
Thus, at least, full NNLO analysis is required to obtain 
the significant evaluation. 
However, the NNLO analysis of the QCD backgrounds 
is still on the way\cite{BINO,TWOL}. 

It has been pointed out that the 
signal-to-background ratio can be improved by considering 
the associated production of the Higgs with a high transverse 
energy {\it jet} ( 
$pp\rightarrow H~jet \rightarrow \gamma\gamma~jet$)\cite{ABD}. 
The existence of a high $P_T$ {\it jet} in the final state allows to 
choose suitable cuts to suppress the QCD background\cite{ABD}. 
In addition, for this process, the ambiguity of the higher order 
corrections is less serious. 
The subprocess with two gluon initial states 
$gg\rightarrow g\gamma\gamma$ again contributes first in the NNLO,  
which is believed to dominate the NNLO contributions. 
However, the LO correction, 
which is dominated by the subprocess 
$qg\rightarrow q\gamma\gamma$, is enough larger than the NNLO 
contributions because of the large gluon  
distribution\cite{FLO}. 
Thus, it is expected that the NLO analysis is fairly accurate 
to understand the QCD background.

To evaluate the NLO corrections of the QCD background, 
we need the tree-level QCD amplitudes for 
six partons ($q\bar{q}\gamma\gamma gg$,$q\bar{q}Q\bar{Q}\gamma\gamma$) 
and one-loop amplitudes  for the five partons($q\bar{q}\gamma\gamma g$). 
In ref.\cite{BILL}, the compact expressions for 
the six-parton tree-level amplitudes are presented. 
The one-loop amplitudes are also discussed in ref.\cite{SIGN,BILL}. 
They provided the systematic procedure to express the  
one-loop five-parton amplitudes with using the known 
results of the $q\bar{q} ggg$ amplitudes or 
"primitive amplitudes"\cite{BERN1}. 
However, the QCD amplitudes include a lot of unnecessary 
terms which disappear in the explicit expression 
of photons amplitudes. 
Thus, their one-loop expressions are still rather complicated.   
In this paper, we carry out the direct calculations of 
the one-loop five-parton amplitudes, which are involving 
two massless quarks, two or three external photons. 
Our final goal of this paper is to obtain the explicit 
expressions for the one-loop $\bar{q}qg\gamma\gamma$ 
amplitudes without using QCD amplitudes.

The paper is organized as follows. 
It is now popular that QCD matrix elements of multi-partons 
are expressed in terms of the color ordered helicity amplitudes. 
In section 2 , we explain some technical prescriptions of these methods. 
We also explain known general forms of tree-level amplitudes with 
external gluons and/or photons. 
In Section 3, we present one-loop helicity amplitudes for 
five-parton processes including two or tree external photons. 
In Section 4, we give some concluding remarks.

%\newpage
%\input chap2.tex
\section{Tree-level Amplitudes}

The color decomposition\cite{color} and helicity basis method\cite{SHB} 
are now standard techniques to express the multi-parton amplitudes.
The color decomposition method is the procedure which 
constructs color ordered gauge invariant partial amplitudes in the QCD. 
At the tree level, amplitudes ${\cal A}_n$ 
involving two quarks and $n-2$ gluons 
can be decomposed into the partial amplitudes 
${\cal M}$'s which are 
characterized by the single string of the group matrices\cite{PARKE1}, 
\[
{\cal A}_n =g^{n-2} \sum_{a_i \in S_{n-2}}
 (T^{a_3}\cdots T^{a_n})
{A}_n(q^{i_1},\bar{q}^{i_2},g^{i_3}\cdots ,g^{i_n}),
\]
where $i_j$ are parton helicities.
$T^a~(a=1,2,\cdots,N^2-1)$ are the matrices of the gauge 
group in the fundamental representation. $S_n$ denotes 
the set of noncyclic permutations over $1,\cdots,n$.  

We also introduce the spinor helicity basis method\cite{SHB}.
We use the popular notations 
for the helicity basis of a massless spinor field $\psi$\cite{PARKE1},
\[
\langle q^{\pm}|
\equiv{1\over2}\overline{\psi}(q)(1\mp \gamma_5),
~~~~~~~~~~~~~~~
|\bar{q}^{\pm}\rangle 
\equiv{1\over2}(1\pm \gamma_5)\psi(\bar{q}).
\]
In this paper, all external momenta are taken to be outgoing.   
All color ordered helicity amplitudes are expressed in 
terms of following spinor products, 
\[
\langle pq\rangle \equiv\langle p^-|q^+\rangle 
,~~~~~
[pq]\equiv\langle p^+|q^-\rangle 
,~~~~~
[pq]\langle qp\rangle
=s_{pq}\equiv 2p\cdot q
.
\]

%koko
We can also describe external gauge fields by using the 
spinor helicity basis. 
The polarization vectors can be written in terms of massless 
spinors 
$|p^\pm\rangle$ and $|k^\pm\rangle$,
\be
\varepsilon^{\pm}(p,k)
=
\pm{\langle p^\pm|\gamma_\mu|k^\pm\rangle 
\over \sqrt{2}\langle k^\mp|p^\pm\rangle },
\label{eq:pol}
\ee
where $p$ is the gauge boson momentum, 
$k$ is the arbitrary momentum which satisfies 
$k^2=0$. 
We call this momentum $k$ as the {\it reference} momentum. 
Physical quantities do not depend on the 
{\it reference} momentum because a change in the reference momentum 
is equivalent to a gauge transformation:
\[
\varepsilon^{+}(p,k)_\mu
\rightarrow
\varepsilon^{+}(p,k')_\mu-\sqrt{2}
{\langle kk'\rangle \over\langle kp\rangle \langle k'p\rangle }p_\mu.
\]
This means that we can choose an appropriate {\it reference} 
momentum for any gauge invariant subset of the full amplitude.
For example, in this formula, we obtain the following identities, 
\be
\slashs{\varepsilon}^{\pm}(p,k)|k^\pm\rangle=0, 
~~~~~~~~~~~~
\langle k^\mp |\slashs{\varepsilon}^{\pm}(p,k)=0.
\label{eq:REFM}
\ee
Using this identities, we can easily show that 
the amplitudes in which all gluons have the same helicity 
vanish\cite{PARKE1}, 
\be
{A}_n^{tree}(\bar{q},q,g_3^+,g_4^+,\cdots,g_n^+)
=0.
\label{eq:MHV}
\ee
From the above formula, we also obtain the simple expression of the 
amplitudes in which one gluon has negative helicity and 
all other gluons have positive helicity
\cite{BILL},
\be
{A}_n^{tree}(\bar{q}_1^+,q_2^-,g_3^+,\cdots,g_j^-,\cdots,g_n^+)
=
i{
\langle 1j \rangle \langle 2 j\rangle^3
\over 
\langle 12 \rangle \langle 23 \rangle 
\cdots \langle n 1 \rangle},
\label{eq:TREE1}
\ee
where the j-th gluon has the negative helicity.
Here, we followed the notations and conventions given 
in ref.\cite{PARKE1}(In section 3, we use different 
normalization condition of group factor $T^a$). 

The amplitudes involving external photons are obtained by 
summing over permutations of gluon matrix elements\cite{PARKE1}. 
The amplitudes containing gluons and photons 
with maximally-helicity-violating(MHV) configuration 
$(-,-,+,\cdots,+)$ are 
\cite{BILL},
\bea
\lefteqn{
{A}_n^{tree}(\bar{q}_1^,q_2^-,g_3,\cdots,g_{r+2},
\gamma_{r+3},\cdots,\gamma_{r+m+2})}
\nonumber\\
&&
~~~~~=
i{
\langle 1i \rangle \langle 2 i\rangle^3
\over 
\langle 12 \rangle \langle 23 \rangle 
\cdots \langle (r+2) 1 \rangle}
\prod_{j=r+3}^{r+m+2}\left[
{\langle21\rangle\over \langle2j\rangle \langle 
j1\rangle}
\right]
,
\label{eq:TREE2}
\eea
where the i-th gluon or photon has the negative helicity. 
To obtain the one-loop level photons amplitudes, 
similar procedure have been applied to the one-loop amplitudes
\cite{SIGN,BILL}. 

The amplitudes with other configuration of helicities 
are obtained by Parity inversion and charge conjugation\cite{BERN1}. 
Parity inversion  reverses the sign of all helicities of 
external legs.  
This conversion is achieved by taking the complex conjugation.  
In terms of the helicity basis method, this operation is 
same as the replacement of spinor products 
$\NB{ij} \leftrightarrow [ji]$, but with no substitution 
of $i \rightarrow -i$. In addition, a factor $-1$ is 
required for each quark anti-quark pair. 
Charge conjugation replaces quarks and anti-quarks 
without changing helicities. From these relations, 
we can reduce the number of independent partial amplitudes. 
For the case of $q\bar{q}g\gamma\gamma$ amplitude, 
we only need to calculate the amplitudes with following three types 
of helicity configurations; 
$(\bar{q}^-, q^+, g^+, \gamma^+, \gamma^+)$ and 
$(\bar{q}^-, q^+, g^\pm, \gamma^\mp, \gamma^+)$. 

%\newpage
%\input chap3.tex
\section{One-loop results}

The one-loop $q\bar{q}\gamma\gamma g$ amplitudes 
can be written by the three types gauge independent 
partial amplitude ${\cal M}_5^i(i=1,2,3)$, 
\be
{\cal A}_5^{1-loop}(q\bar{q}\gamma\gamma g)
=-{e_q^2g^3T^a\over 2}\left\{
{1\over N_c}{\cal M}_5^1(q\bar{q}\gamma\gamma g)
+N_c {\cal M}_5^2(q\bar{q}\gamma\gamma g)\right\}
-
\sum_{i=1}^{n_f}
{e_{q_i}^2g^3T^a \over 2}
{\cal M}_5^3(q\bar{q}\gamma\gamma g).
\label{FPA}
\ee
${\cal M}_5^1$ and ${\cal M}_5^2$ 
have different dependence on the color factor  
$N_c(=3 ~\mbox{for QCD})$  
and ${\cal M}_5^3$ is the fermion loop contribution. 
$e_q$ is the charge of the external quarks. 
$e_{q_i}$ are the charges of loop fermions and $n_f$ is flavor. 
To explain the one-loop results, we change normalization condition 
of the group generators as $Tr(T^aT^b)=\delta^{ab}/2$. 

To carry out the one-loop calculations, 
we need the information on the Feynman integrals. 
We follow the technology of the Feynman integral calculation 
which discussed in ref.\cite{NEE,BDK6}. 
We use dimensionally regulated one-loop integrals 
in $4-2\epsilon$ dimensions.  
In the conventional dimensional regularization(CDR) scheme, 
both momentum components and helicity states are dealt with 
in $D~(=4-2\epsilon)$ dimensions\cite{BDK2}. 
Thus, all gluons and photons have $2-2\epsilon$ helicity states. 
On the other hand, the spinor helicity basis is defined in four 
dimensions. To apply the helicity basis method to the one-loop 
calculations, we need some modification in the regularization scheme.
The 't Hooft and Veltman scheme is one of the solutions in which  
all polarization of observed particles are dealt with in four 
dimensions (then observed gluons and photons have 2 helicity state). 
The four-dimensional helicity (FDH) scheme \cite{BDK2} 
is a more efficient approach when using the spinor helicity method. 
In this scheme 
momentum components of unobserved particles are dealt with $4-2\epsilon$ 
dimensions, but all helicities are treated in four dimensions. 
Thus, both observed and unobserved gauge bosons have 2 helicity states. 

\subsection{${\cal M}_5^1(\bar{q}qg\gamma\gamma)$ }

\begin{figure*}
\begin{center}
\resizebox{1.5cm}{2cm}{\includegraphics{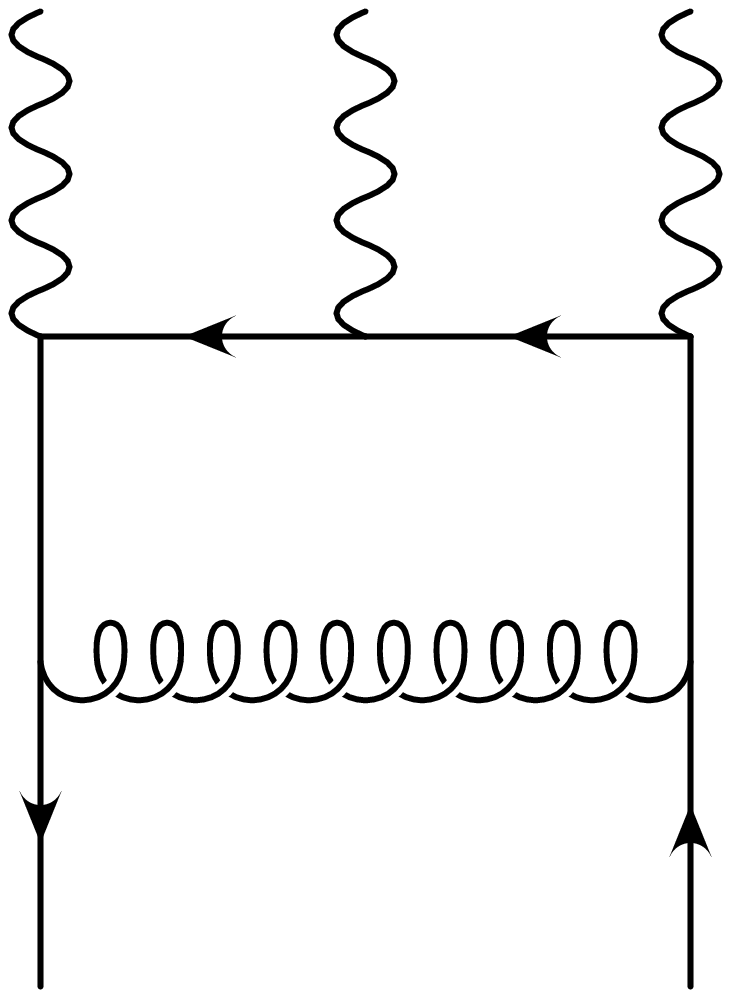} }
\resizebox{1.5cm}{2cm}{\includegraphics{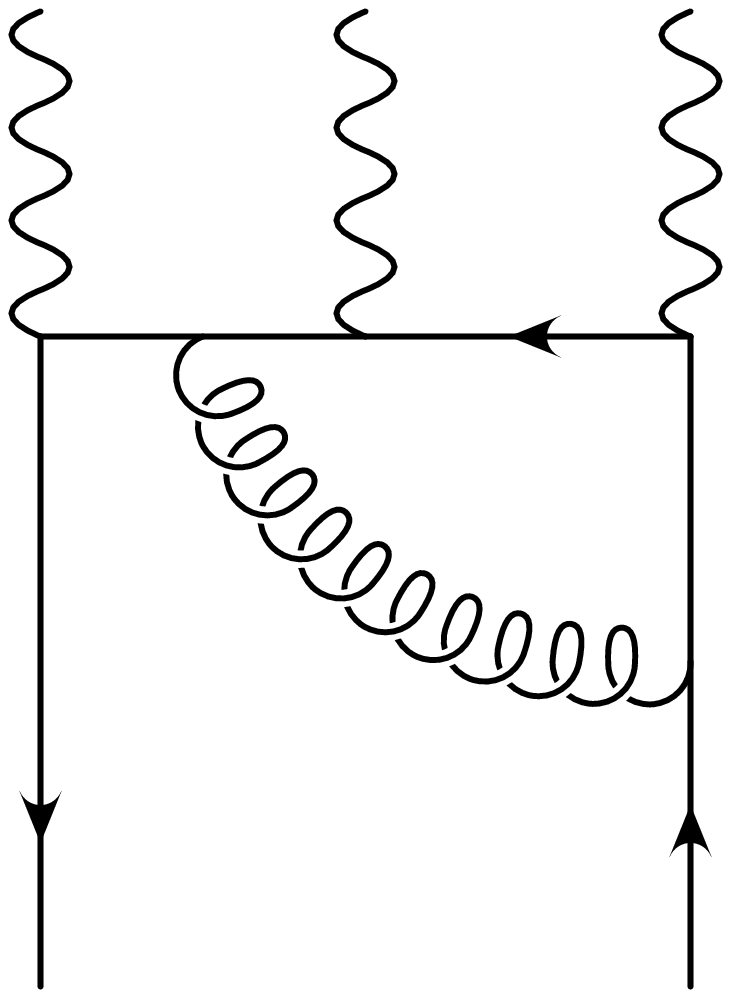} }
\resizebox{1.5cm}{2cm}{\includegraphics{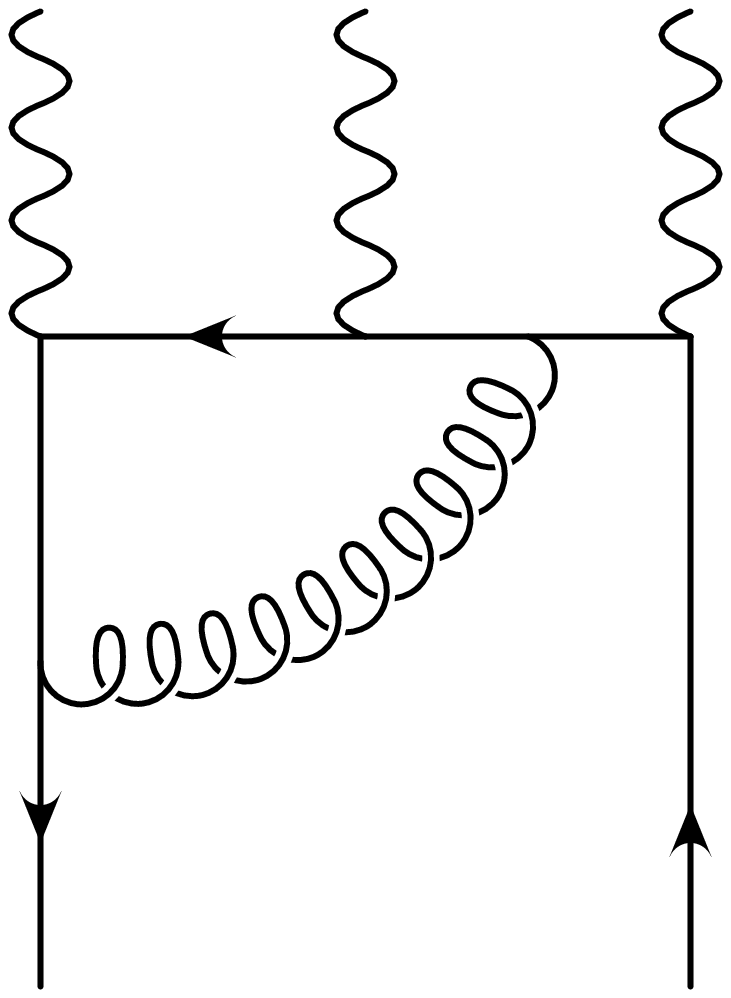} }
\resizebox{1.5cm}{2cm}{\includegraphics{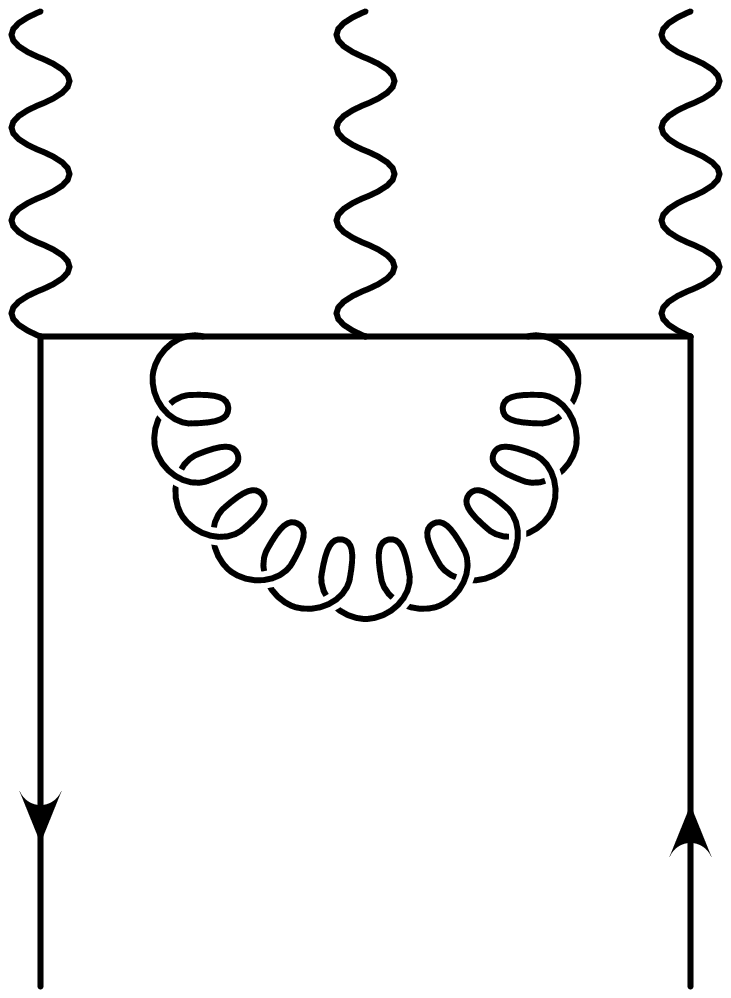} }
\resizebox{1.5cm}{2cm}{\includegraphics{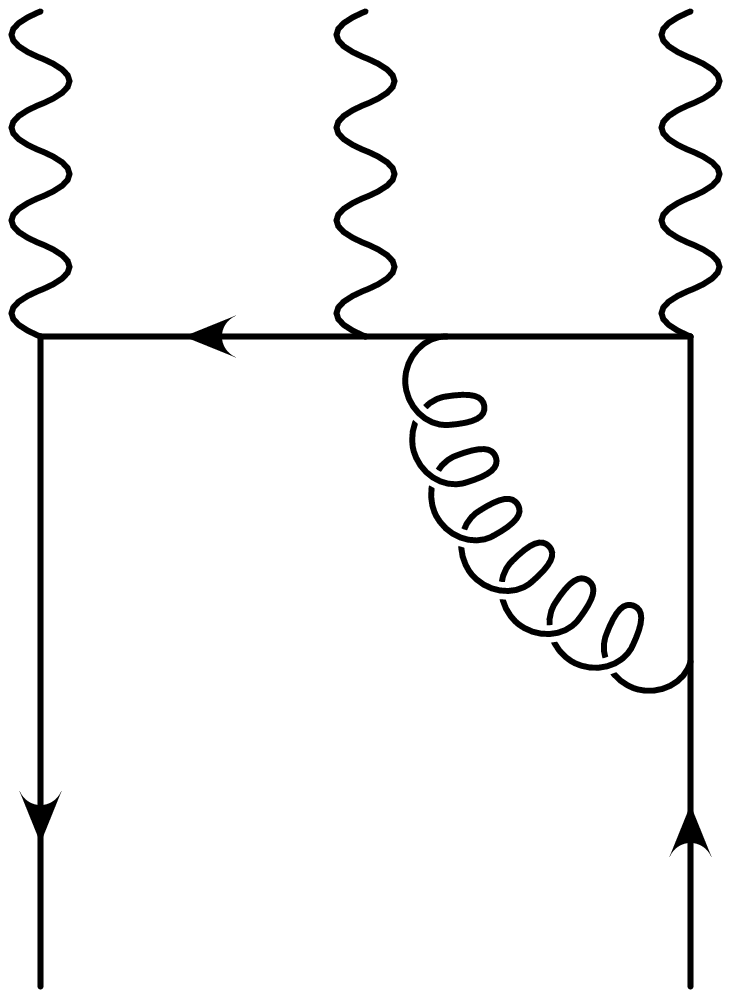} }
\resizebox{1.5cm}{2cm}{\includegraphics{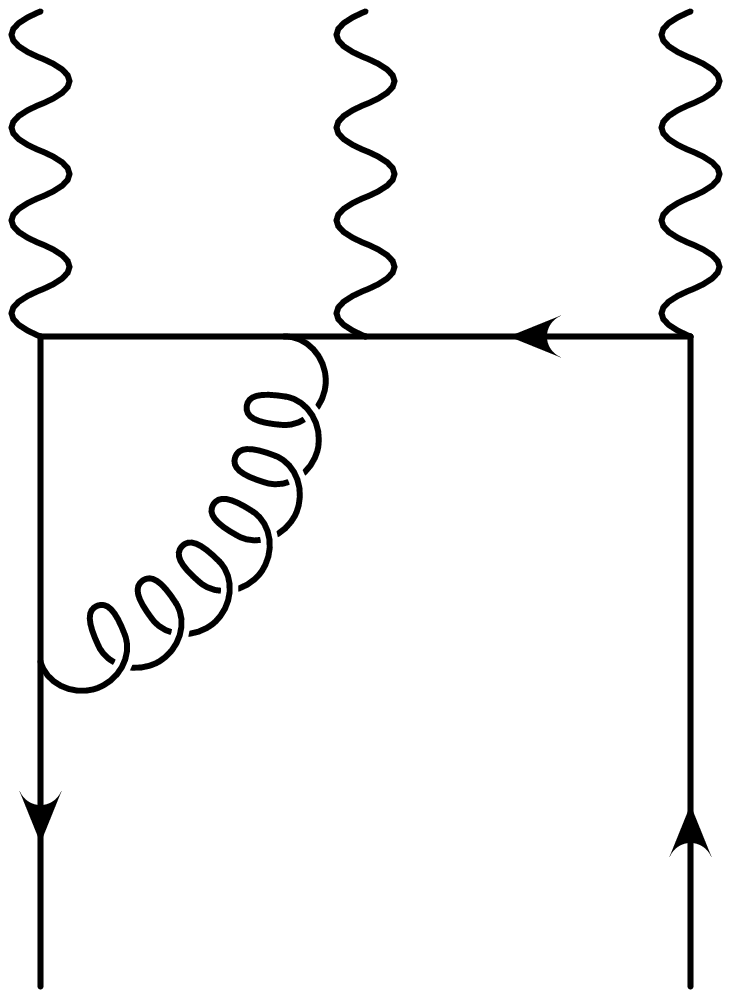} }
\resizebox{1.5cm}{2cm}{\includegraphics{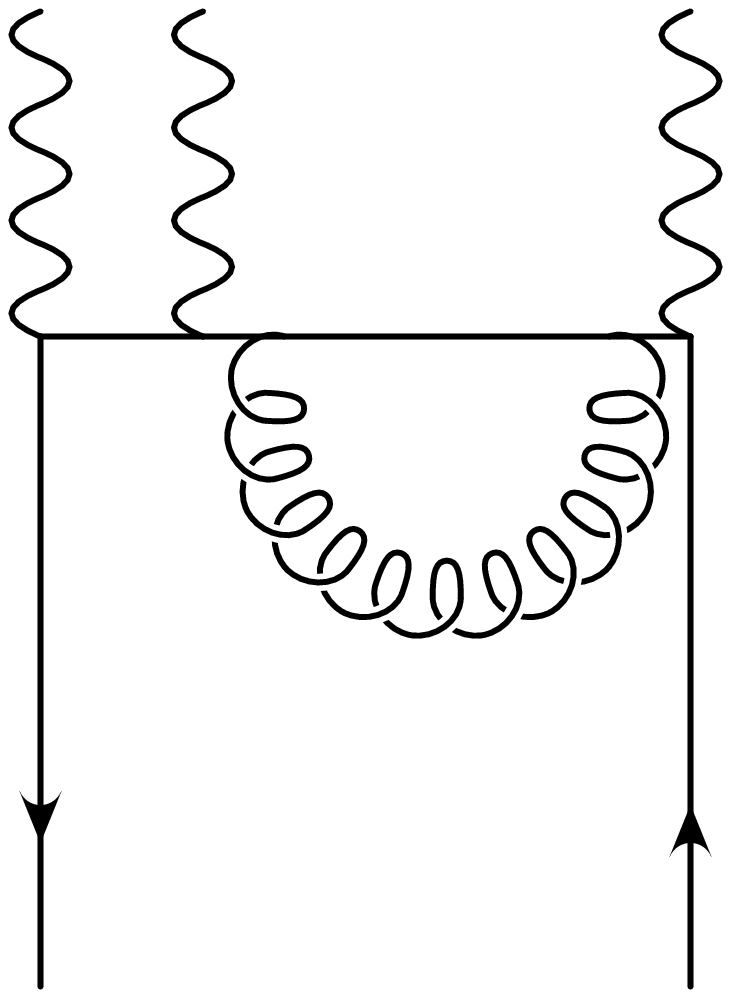} }
\resizebox{1.5cm}{2cm}{\includegraphics{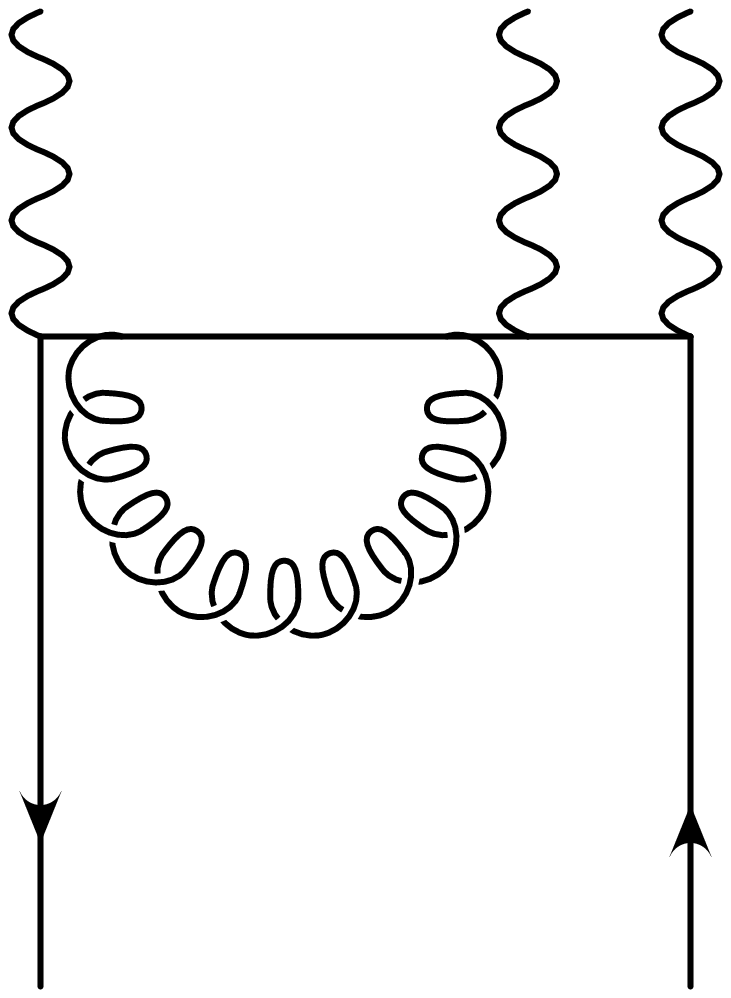} }
\end{center}
\caption{
Feynman diagrams for ${\cal O}(1/N_c)$ contributions. 
%One of the external legs is a gluon and others are photons.
}
\label{QEDfig}
\end{figure*}

The Feynman diagrams which contribute to 
the partial amplitudes ${\cal M}_5^1$ are given in figure \ref{QEDfig}. 
Here, one of the external gauge boson legs 
is a gluon and others are photons. 
We notice that 
the amplitudes ${\cal M}_5^1$ are obtained from the 
two-quark three-photon amplitudes 
${\cal A}(q\bar{q}\gamma\gamma\gamma)$ 
by converting one of the photons into a gluon. 
\bea
{\cal M}_5^1
(q,\bar{q},g,\gamma,\gamma)
&=&
{m}_5
(q,\bar{q},\gamma\rightarrow g,\gamma,\gamma), 
\nonumber\\
%&&\nonumber\\
{\cal A}^{1-loop}
(q,\bar{q},\gamma,\gamma,\gamma)
&=&
e_q^3g^2{N_c^2-1\over 2N_c}
{m}_5
(q,\bar{q},\gamma,\gamma,\gamma).
\nonumber
\eea
Therefore, we present  
the $q\bar{q}\gamma\gamma\gamma$ amplitudes at first.

From the analogy of eq.(\ref{eq:REFM}) and eq.(\ref{eq:MHV}), 
the tree level amplitudes in which all of photons have same helicity 
vanish, 
\[
{m}_n^{tree}(\bar{q},q,\gamma_3^+,\gamma_4^+,\cdots,\gamma_n^+)
=0.
\]
This result ensure that the corresponding amplitudes at one-loop level 
must be infrared and ultraviolet finite. 
The explicit form of the one-loop amplitude for the helicity 
$({q}_1^+,\bar{q}_2^-,\gamma_3^+,\gamma_4^+,\gamma_5^+)$ 
is, 
\bea
\lefteqn{
{m}_5
({q}_1^+ \bar{q}_2^-  \gamma_3^+\gamma_4^+\gamma_5^+)%}
%\nonumber\\
%&=&
=
{i \over (4\pi)^2}
{\sqrt{2}
\over 
[24]
\NB{14}\NB{35}
\NB{45}\NB{34}
}}
\nonumber\\
&\times&\Biggl[
{\frac {\left (\PB{{51}}\PB{{23}}\NB{{13}}\NB{{25}}+\NB{{34}}\NB{{12}}\PB{{24
}}\PB{{13}}\right )s_{{13}}\left (s_{{12}}+s_{{25}}\right )}{s_{{24}}s_
{{51}}}}
\nonumber\\
&~&
+{\frac {\left (\PB{{51}}\PB{{23}}\NB{{13}}\NB{{25}}-\NB{{23}}\PB{{13}}\PB{{24
}}\NB{{14}}+2\,s_{{25}}s_{{13}}\right )s_{{23}}}{s_{{24}}}}
\nonumber\\
&~&
+{\frac {\left (-\PB{{23}}\NB{{34}}\PB{{45}}\NB{{25}}-\NB{{45}}\PB{{51}}\NB{{
12}}\PB{{24}}\right )s_{{45}}\left (s_{{12}}+s_{{25}}\right )}{s_{{51}}
s_{{23}}}}
\nonumber\\
&~&
+{\frac {\left (-2\,\PB{{51}}\NB{{12}}\PB{{23}}\NB{{35}}+2\,\PB{{45}}\PB{{23}
}\NB{{35}}\NB{{24}}+s_{{45}}s_{{12}}\right )s_{{24}}}{s_{{23}}}}
\nonumber\\
&~&
-{\frac {\left (-2\,\PB{{51}}\NB{{12}}\PB{{23}}\NB{{35}}-2\,\PB{{34}}\NB{{45
}}\PB{{51}}\NB{{13}}+s_{{34}}s_{{12}}-{s_{{34}}}^{2}\right )s_{{23}}}{s_
{{51}}}}
\nonumber\\
&~&
+{\frac {s_{{34}}\left (-{s_{{12}}}^{2}+s_{{34}}s_{{12}}+s_{{34}}s_{{13
}}\right )}{s_{{51}}}}
\nonumber\\
&~&
+\PB{{45}}\PB{{12}}\NB{{25}}\NB{{14}}+\NB{{45}}\PB{{51}}\NB{{12}}\PB{{24}}+2\,{
s_{{45}}}^{2}+2\,s_{{45}}s_{{34}}
\nonumber\\
&~&
+s_{{51}}\left (s_{{51}}-3\,s_{{45}}+s_{{34}}+5\,s_{{12}}\right )
\biggr]+(3 \leftrightarrow 5).
\nonumber
\eea 

An independent non-vanishing tree-level partial amplitude 
$m_5^{tree}$ is given by the helicity 
configuration  
(${q}^-,\bar{q}^+,\gamma^-,\gamma^+,\gamma^+$), 
\be
{m}_5^{\mbox{tree}}
({q}_1^+ \bar{q}_2^-  \gamma_3^+\gamma_4^-\gamma_5^+) 
={i 2\sqrt{2}
\NB{12}\NB{24}^2
\over 
\NB{25}\NB{23}
\NB{51}\NB{13}.
}
\label{eq:NONSIN}
\ee
Here, we define the partial amplitudes $m_5$'s as,
\[
 A_5^{tree}(q\bar{q}\gamma\gamma\gamma)
\equiv 
 e^3m_5^{tree}(q\bar{q}\gamma\gamma\gamma)
\]
The difference between eq.(\ref{eq:TREE2}) and 
eq.(\ref{eq:NONSIN}) 
in the factor $2\sqrt{2}$ comes from the different 
normalization of group factor $T^a$. 
The corresponding one-loop level partial amplitude 
possesses ultraviolet and infrared divergences. 
It is well known that the singular part of 
one-loop n-point amplitudes have 
the universal structure\cite{KUN,BC},
\be
{m}^{{1-loop}}_n |_{\mbox{singular}}
=
c_\Gamma {m}^{\mbox{tree}}_n 
\left[
-{1\over \epsilon^2}\sum_{j=1}^n {\cal S}_j^{[n]}
\left({\mu^2\over -s_{j,j+1}}\right)^\epsilon
+{\cal C}^{[n]}{1\over \epsilon}\right],
\label{eq:SIN}
\ee
where
\[
c_\Gamma\equiv {(4\pi)^\epsilon\over 16\pi^2}
{\Gamma(1+\epsilon)\Gamma^2(1-\epsilon)
\over
\Gamma(1-2\epsilon)}.
\]
with $D=4-2\epsilon$ and $\mu$ is the renormalization scale.
${m}^{tree}$ is the tree-level partial amplitude. 
${\cal S}_j^{[n]}$ are the coefficients of 
the soft singularities and 
${\cal C}_j^{[n]}$ are the sum of the coefficients 
of the collinear and 
ultraviolet singularities which depend on the 
particle contents of amplitudes. 
For the massless QCD, accurate values of 
the coefficients ${\cal S}_j^{[n]}$ 
and ${\cal C}_j^{[n]}$ are well known\cite{BERN1,BC}. 
Because of this fact, it is easy to check the singular part 
of one-loop level amplitudes. 
We will mention this matter in the end of this section. 
Here, we present the amplitudes in which
ultraviolet divergences are unsubtracted. 

We also introduce following functions 
to explain the non-singular parts, 
\bea
L(s,t,u)&\equiv& 
Li_2\left(1-{s\over u}\right)
+Li_2\left(1-{t\over u}\right)
+ln{s\over u}ln{t\over u}
-{\pi^2\over6}
\nonumber\\
Rln^n\left({s\over t}\right)
&\equiv&
{ln\left({s\over t}\right)\over (s-t)^n}, 
\nonumber
\eea 
where $Li_2(Z)$ is the dilogarithm function\cite{DILOG}:
\[
Li_2(Z)\equiv -\int_0^Z{dx\over x}log(1-x).
\]
The explicit form of the one-loop partial amplitude 
for the helicity configuration  
(${q}^-,\bar{q}^+,\gamma^-,\gamma^+,\gamma^+$) 
is,
\bea
\lefteqn{
{m}_5
({q}_1^+ \bar{q}_2^-  \gamma_3^+\gamma_4^-\gamma_5^+)}
\nonumber\\
&=&
%{4\over i(4\pi)^2}
c_\Gamma {m}^{\mbox{tree}} 
\left\{
-{2\over \epsilon^2} 
\left({\mu^2\over -s_{12}}\right)^\epsilon
-{3\over \epsilon(1-2\epsilon)}
\left({\mu^2\over -s_{12}}\right)^\epsilon 
-3 -\delta_D\right\}
\nonumber\\
&&+
{i \over (4\pi)^2}
{\sqrt{2}\over \PB{24}^2\NB{35}^2\NB{12}}
\Biggl[
f_1~ln\left({s_{34}\over s_{12}}\right)
+f_2~ln\left({s_{14}\over s_{45}}\right)
+{f_3~Rln^1\left({s_{23}\over s_{45}}\right)}
+{f_4~Rln^2\left({s_{14}\over s_{23}}\right)}
\nonumber\\
&&
-4\,{\frac {\NB{{35}}\PB{{24}}s_{{24}}{\NB{{12}}}^{2}s_{{12}}\left (-\PB{{
13}}\NB{{34}}+\PB{{51}}\NB{{45}}\right )}{\NB{{23}}\NB{{51}}{s_{{13}}}^{2}}
}\tilde{L}(s_{23},s_{12},s_{45})
\nonumber\\
&&
-4\,{\frac {{\NB{{45}}}^{2}{\NB{{12}}}^{3}{\PB{{24}}}^{2}\NB{{35}}}{{\NB{{
51}}}^{3}\NB{{23}}}}
\left(
\tilde{L}(s_{14},s_{45},s_{23})
+{s_{51}^2\over s_{13}^2}\tilde{L}(s_{23},s_{12},s_{45})\right)
\nonumber\\
&&
- 4\,{\frac {\NB{{12}}\NB{{35}}{s_{{24}}}^{2}}{\NB{{51}}\NB{{23}}}}
(L(s_{12},s_{51},s_{34})+L(s_{34},s_{24},s_{51}))
\nonumber\\
&&+
\Biggl\{
2\,{\frac {\PB{{35}}\left (-\PB{{14}}\NB{{51}}\NB{{34}}+\NB{{23}}\PB{{24}}\NB
{{45}}\right )\left (s_{{34}}+s_{{45}}\right )}{s_{{35}}}}
-2\,{\frac {{s_{{34}}}^{2}\left (-s_{{12}}+s_{{34}}-2\,s_{{51}}\right 
)}{s_{{51}}}}
\nonumber\\
&&
-2\,{\frac {\PB{{23}}\left (\PB{{51}}\NB{{12}}\NB{{35}}-\NB{{34}}\PB{{45}}\NB
{{25}}\right )\left (s_{{51}}-s_{{34}}\right )^{2}}{s_{{23}}s_{{51}}}}
-2\,\NB{{34}}\PB{{45}}\NB{{51}}\PB{{13}}+2\,\NB{{23}}\PB{{24}}\NB{{45}}\PB{{35
}}
\nonumber\\
&&
+2\,s_{{23}}s_{{12}}
-4\,s_{{12}}\left (s_{{51}}-s_{{34}}\right )
\Biggr\}L(s_{45},s_{34},s_{12})
%\nonumber\\
%&&
~+~R_0~
 ~+~ (3\leftrightarrow 5)\Biggl].
\nonumber
\eea
The scheme dependence is controlled by the parameter $\delta_D$. 
We obtain $\delta_D=0$ in the FDH scheme and 
$\delta_D=1$ in the 't Hoot-Veltman scheme. 
Here, we also used the following definitions, 
\bea
\tilde{L}(s_{23},s_{12},s_{45})
&\equiv&
{L}(s_{23},s_{12},s_{45})
-{s_{13}}Rln^1\left({s_{23} \over s_{45}}\right)
\nonumber\\
\tilde{L}(s_{14},s_{45},s_{23})
&\equiv&
{L}(s_{14},s_{45},s_{23})
-{s_{51}}Rln^1\left({s_{23}\over s_{45}}\right)
-{s_{51}}Rln^1\left({s_{14}\over s_{23}}\right)
\nonumber
\eea
$f_i~(i=1,\cdots,4)$ and $R_0$ are given by, 
\bea
f_1&=&
-{\frac {\PB{{23}}\left (-\NB{{13}}\NB{{25}}\PB{{51}}+\NB{{24}}\PB{{45}
}\NB{{35}}\right )\left (\,s_{{25}}+3\,s_{{13}}-3\,s_{{45}}\right )
}{s_{{23}}}}
\nonumber\\
&~&
+2\,{\frac {\PB{{51}}\left (\NB{{45}}\NB{{12}}\PB{{24}}-\NB{{13}}\PB{{23}}
\NB{{25}}\right )s_{{24}}}{s_{{51}}}}
%\nonumber\\
%&~&
-3\,{\frac {\left (\PB{{45}}\NB{{14}}\PB{{13}}\NB{{35}}+\NB{{45}}\PB{{14}}\NB
{{12}}\PB{{25}}\right ){s_{{45}}}^{2}}{s_{{25}}s_{{13}}}}
\nonumber\\
&~&
-2\,{\frac {\left (\NB{{51}}\PB{{25}}\NB{{23}}\PB{{13}}+\NB{{45}}\PB{{51}}\NB
{{12}}\PB{{24}}-s_{{51}}s_{{45}}\right )\left (s_{{45}}+s_{{23}}\right 
)}{s_{{25}}}}
\nonumber\\
&~&
+{\frac {\left (\NB{{45}}\PB{{14}}\NB{{12}}\PB{{25}}-\NB{{51}}\PB{{12}}\NB{{23
}}\PB{{35}}\right )s_{{45}}}{{s_{25}}}}
%\nonumber\\
%&~&
-{\frac {s_{{45}}\left (s_{{45}}+s_{{23}}\right )\left (s_{{51}}-2\,s_
{{12}}\right )}{s_{{25}}}}
+{\frac {{s_{{45}}}^{2}\left (2\,s_{{51}}+3
\,s_{{45}}\right )}{s_{{25}}}}
\nonumber\\
& &
-{\frac {\PB{{13}}\left (\PB{{45}}\NB{{14}}\NB{{35}}-\NB{{51}}\PB{{25}}\NB{{
23}}\right )\left (s_{{25}}+4\,s_{{45}}\right )}{s_{{13}}}}
%\nonumber\\
%&~&
+3\,{\frac {{s_{{45}}}^{2}\left (s_{{23}}-s_{{45}}\right )}{s_{{13}}}}
\nonumber\\
&&
-5\,\NB{{51}}\PB{{12}}\NB{{23}}\PB{{35}}+5\,\PB{{51}}\NB{{25}}\PB{{23}}\NB{{13
}}
+2\,{s_{{51}}}^{2}+{s_{{34}}}^{2}-3\,s_{{34}}s_{{51}}-2\,s_{{23}}s_{{45
}}-2\,{s_{{12}}}^{2}
\nonumber\\
&&
-3\,s_{{51}}s_{{12}}+3\,s_{{34}}s_{{45}}+4\,s_{{34
}}s_{{23}}+s_{{51}}s_{{45}}-s_{{12}}s_{{34}}+s_{{12}}s_{{45}}+7\,{s_{{
45}}}^{2}+4\,{s_{{23}}}^{2} 
%\nonumber
%\eea
\nonumber\\ &&
\nonumber\\ 
%\bea
f_2&=&
-4\,{\frac {\PB{{25}}\left
(-\NB{{12}}\PB{{14}}\NB{{45}}+\NB{{51}}\NB{{23}}
\PB{{13}}\right )\left (s_{{23}}+3\,s_{{12}}\right )}{s_{{25}}}}
\nonumber\\
&&+
8\,{\frac {\PB{{23}}\left (\PB{{51}}\NB{{12}}\NB{{35}}-\NB{{25}}
\PB{{45}}\NB{
{34}}\right )s_{{12}}}{s_{{23}}}}
\nonumber\\
&&
-4\,s_{{45}}s_{{12}}+4\,{s_{{23}}}^{2}+8\,{s_{{12}}}^{2}-4\,{s_{{34}}}
^{2}+16\,s_{{12}}s_{{23}}-8\,s_{{34}}s_{{45}}
%\nonumber
%\eea
\nonumber\\ &&
\nonumber\\ 
%\bea
f_3&=&
{\frac {\left \PB{23}
(\NB{{12}}\PB{{51}}\NB{{35}}-\NB{{34}}\PB{{45}}\NB{{25}}\right )
{s_{{34}}}^{2}\left (-3\,s_{{45}}+2\,s_{{12}}\right 
)}{s_{{51}}s_{{23}}}}
\nonumber\\
&&
-{\frac {\left (\PB{{51}}\NB{{12}}\PB{{23}}\NB{{35}}+\PB{{34}}\NB{{45}}\PB{{
51}}\NB{{13}}+{s_{{34}}}^{2}-s_{{34}}s_{{12}}\right )
\left
(s_{{23}}s_{12}+s_{{23}}s_{{13}}+s_{{34}}s_{{12}}+3\,s_{{34}}s_{{13}}
\right )}{s_{{51}}}}
\nonumber\\
&&
-{\frac {\left (\PB{{34}}\NB{{45}}\PB{{51}}\NB{{13}}-\PB{{23}}\NB{{34}}\PB{{
45}}\NB{{25}}\right )s_{{34}}\left (s_{{45}}-3\,s_{{34}}-s_{{23}}
\right )}{s_{{51}}}}
\nonumber\\
&&
+{\frac {\PB{{23}}\left (-\PB{{51}}\NB{{12}}\NB{{35}}+\NB{{25}}\PB{{45}}\NB{{
34}}\right )\left (
2s_{51}s_{12}-4s_{34}s_{12}+5s_{34}s_{45}-2s_{45}s_{51}
\right )}{s_{{23}}}}
\nonumber\\
&&
-3\,\left (\PB{{34}}\NB{{45}}\PB{{51}}\NB{{13}}-\PB{{23}}\NB{{34}}\PB{{45}}\NB
{{25}}+s_{{45}}s_{{13}}+s_{{34}}s_{{12}}+s_{{45}}s_{{23}}\right )
s_{{34}}
\nonumber\\
&&
-2\,\NB{{34}}\left (\NB{{12}}\PB{{23}}\PB{{14}}-\PB{{45}}\NB{{51}}\PB{{13}}
\right )\left (-s_{{45}}+s_{{12}}\right )
%\nonumber\\
%&&
-2\,s_{{45}}\left (-{s_{{34}}}^{2}+s_{{51}}s_{{12}}-s_{{23}}s_{{34}}
\right )
\nonumber\\
&&           
+2\,\left (s_{{51}}-s_{{34}}\right ){s_{{12}}}^{2}
%\nonumber\\
%&&
-5\,s_{{34}}s_{{45}}\left (s_{{45}}-s_{{34}}+s_{{51}}\right )
%\nonumber\\
%&&
+5\,s_{{34}}s_{{23}}\left (s_{{12}}+s_{{45}}\right )
\nonumber\\
&&
-2\,{s_{{34}}}^{2}\left (s_{{23}}-s_{{45}}\right )
%\nonumber\\
%&&
+s_{{34}}\left (s_{{51}}s_{{12}}+s_{{45}}s_{{23}}\right )
%\nonumber
%\eea
\nonumber\\ &&
\nonumber\\
%\bea
f_4&=&
{\frac {\PB{{23}}\left (\PB{{51}}\NB{{12}}\NB{{35}}-\NB{{25}}\PB{{45}}
\NB{{34}}\right )s_{34}
({s_{{34}}}-s_{51})\left (s_{{51}}^2-s_{{45}}^2\right )
}{s_{{51}}s_{{23}}}}
+{\frac {{s_{{34}}}^{2}{s_{{45}}}^{2}\left (s_{{12}}-s_{{34}}\right )}{
s_{{51}}}}
\nonumber\\
&&
-{\frac {\PB{{51}}\left (\NB{{35}}\PB{{23}}\NB{{12}}+\PB{{34}}\NB{{45}}\NB{{
13}}\right )\left (s_{{12}}-s_{{34}}\right )\left (s_{{12}}s_{{23}}-s_
{{23}}s_{{34}}+2\,s_{{34}}s_{{45}}\right )}{s_{{51}}}}
\nonumber\\
&&
-{\frac {2~\PB{{23}}\left (\PB{{51}}\NB{{12}}\NB{{35}}-\NB{{25}}\PB{{45}}\NB{{
34}}\right )s_{45}
\left (s_{{51}}-s_{{34}}\right )
\left (s_{{51}}+s_{{45}}\right )}{s_{{23}}}}
\nonumber\\
&&
+
2\,\left (-\PB{{12}}\NB{{23}}\PB{{34}}\NB{{14}}+\NB{{12}}\PB{{13}}\NB{{34}}\PB
{{24}}+s_{{45}}s_{{51}}+{s_{{34}}}^{2}\right )s_{{45}}\left (s_{{51}}+
s_{{45}}\right )
\nonumber\\
&&
-2\,\left (2\,\PB{{34}}\NB{{45}}\PB{{51}}\NB{{13}}+s_{{34}}s_{{51}}-s_{{45
}}s_{{14}}\right )s_{{34}}\left (-s_{{34}}-s_{{23}}+s_{{12}}\right )
\nonumber\\
&&
-2\,\left (\PB{{34}}\PB{{51}}\NB{{14}}\NB{{35}}+\NB{{45}}\NB{{12}}\PB{{14}}\PB
{{25}}+{s_{{12}}}^{2}-s_{{34}}s_{{12}}\right )\left (-s_{{45}}+s_{{12}
}\right )s_{{23}}
\nonumber\\
&&
+\left (\NB{{45}}\NB{{12}}\PB{{14}}\PB{{25}}+\NB{{23}}\PB{{34}}\NB{{45}}\PB{{25
}}-s_{{23}}s_{{34}}-s_{{45}}s_{{12}}\right )\left (2\,s_{{12}}+s_{{23}
}\right )s_{{51}}
\nonumber\\
&&
+\left (2\,\NB{{23}}\PB{{34}}\NB{{45}}\PB{{25}}+s_{{12}}s_{{23}}-2\,s_{{34}
}s_{{45}}\right )s_{{12}}\left (s_{{12}}-s_{{34}}+s_{{45}}\right )
\nonumber\\
&&
-{s_{{45}}}^{2}\left (s_{{45}}s_{{23}}+3\,s_{{12}}s_{{23}}-2\,{s_{{34}
}}^{2}-s_{{34}}s_{{12}}\right )
%\nonumber\\
%&&
-s_{{12}}s_{{23}}\left (-s_{{34}}s_{{12}}+2\,s_{{45}}s_{{51}}\right )
\nonumber\\
&&
-{s_{{34}}}^{2}\left (s_{{34}}s_{{51}}+s_{{12}}s_{{23}}-4\,s_{{45}}s_{
{51}}-s_{{51}}s_{{12}}-s_{{45}}s_{{23}}+2\,s_{{51}}s_{{23}}\right )
\nonumber\\
&&
-s_{{12}}\left (-s_{{34}}+2\,s_{{45}}+s_{{23}}\right ){s_{{51}}}^{2}
%\nonumber\\
%&&
+s_{{23}}s_{{34}}\left (2\,s_{{45}}s_{{23}}+s_{{51}}s_{{23}}+5\,s_{{45}
}s_{{51}}\right )
%\nonumber
%\eea
\nonumber\\ &&
\nonumber\\
%\bea
R_0&=&
{\frac {\PB{{25}}\left (\NB{{45}}\PB{{34}}\NB{{23}}+\NB{{51}}\NB{{24}}\PB{{14
}}\right )\left (2\,s_{{51}}-3\,s_{{34}}-s_{{45}}\right )s_{{24}}}{
\left (s_{{34}}+s_{{45}}\right )s_{{25}}}}
\nonumber\\
&&
+{\frac {\NB{{14}}\PB{{34}}\left (-\PB{{12}}\NB{{23}}+\PB{{51}}\NB{{35}}
\right )s_{{35}}s_{{12}}}{s_{{14}}s_{{23}}}}
%\nonumber\\
%&&
+{\frac {{s_{{35}}}^{2}\left (s_{{23}}+s_{{34}}\right )}{s_{{51}}}}
\nonumber\\
&&
-{\frac {\NB{{45}}\PB{{24}}\left (-\PB{{51}}\NB{{12}}+\PB{{35}}\NB{{23}}
\right ){s_{{35}}}^{2}}{s_{{51}}\left (s_{{51}}+s_{{45}}\right )}}
%\nonumber\\
%&&
-{\frac {{s_{{35}}}^{2}\left (s_{{12}}+s_{{34}}+s_{{45}}+s_{{23}}
\right )}{s_{{51}}+s_{{45}}}}
\nonumber\\
&&
+{\frac {\left
(-\NB{{51}}\PB{{12}}\NB{{23}}\PB{{35}}+\PB{{34}}\PB{{51}}
\NB{{14}}\NB{{35}}-s_{{51}}s_{{23}}\right )s_{{12}}s_{{35}}}
{s_{{23}}\left (s_{{51}}+s_{{45}}\right )}}
\nonumber\\
&&
+2\,{\frac {s_{{12}}s_{{14}}\left
(\PB{{51}}\NB{{12}}\PB{{23}}\NB{{35}}
-\NB{{34}}\PB{{45}}\NB{{51}}\PB{{13}}+{s_{{45}}}^{2}
-s_{23}s_{13}-2 s_{23}s_{45}
\right )}{s_{{23}}
\left (s_{{34}}+s_{{45}}\right )}}
\nonumber\\
&&
+2\,{\frac {\left
(\NB{{51}}\PB{{25}}\NB{{23}}\PB{{13}}-\PB{{34}}\NB{{45}}
\PB{{51}}\NB{{13}}+{s_{{51}}}^{2}
+s_{45}s_{51}+s_{45}^2
\right )s_{{14}}}{s_{{34}}+s_{{45}}}}
\nonumber\\
&&
+3\,{\frac {\PB{{51}}\left (\NB{{13}}\NB{{45}}\PB{{34}}+\NB{{12}}\PB{{23}}\NB{
{35}}\right )\left (s_{{23}}+s_{{34}}\right )^{2}}{s_{{51}}s_{{23}}}}
%\nonumber\\
%&&
-3\,{\frac {\left (s_{{12}}-s_{{45}}\right ){s_{{34}}}^{2}}{s_{{23}}}}
\nonumber\\
&&
+{\frac {\PB{{23}}\left (\NB{{25}}\PB{{45}}\NB{{34}}-\PB{{51}}\NB{{12}}\NB{{
35}}\right )\left (2\,s_{{25}}+3\,s_{{34}}\right )}{s_{{23
}}}}
%\nonumber\\
%&&
+2\,{\frac {s_{{45}}s_{{12}}\left (s_{{51}}+s_{{45}}\right )}{s_{{23}}}
}
%\nonumber\\
%&&
%+2\,{\frac {s_{{14}}\left (s_{{45}}s_{{51}}-s_{{12}}s_{{13}}
%+{s_{{45}}}^{2}-2\,s_{{45}}s_{{12}}\right )}{s_{{34}}+s_{{45}}}}
\nonumber\\
&&
+14\,\NB{{23}}\PB{{34}}\NB{{45}}\PB{{25}}+7\,\PB{{12}}\NB{{23}}\PB{{34}}\NB{{14
}}-7\,\PB{{51}}\NB{{12}}\PB{{23}}\NB{{35}}+7\,s_{{45}}s_{{34}}+5\,{s_{{51}
}}^{2}
\nonumber\\
&&
-6\,s_{{51}}s_{{23}}+{s_{{23}}}^{2}-3\,s_{{51}}s_{{12}}+s_{{12}}
s_{{23}}-s_{{23}}s_{{45}}-5\,{s_{{34}}}^{2}+2\,{s_{{45}}}^{2}+{s_{{12}
}}^{2}-4\,s_{{23}}s_{{34}}
\nonumber\\
&&
+3\,s_{{45}}s_{{51}}+4\,s_{{12}}s_{{34}}-7\,
s_{{45}}s_{{12}}-s_{{34}}s_{{51}}
.
\nonumber
\eea

\subsection{${\cal M}_5^2(\bar{q}qg\gamma\gamma)$}

\begin{figure}[b]
\begin{center}

\resizebox{1.3cm}{2cm}{\includegraphics{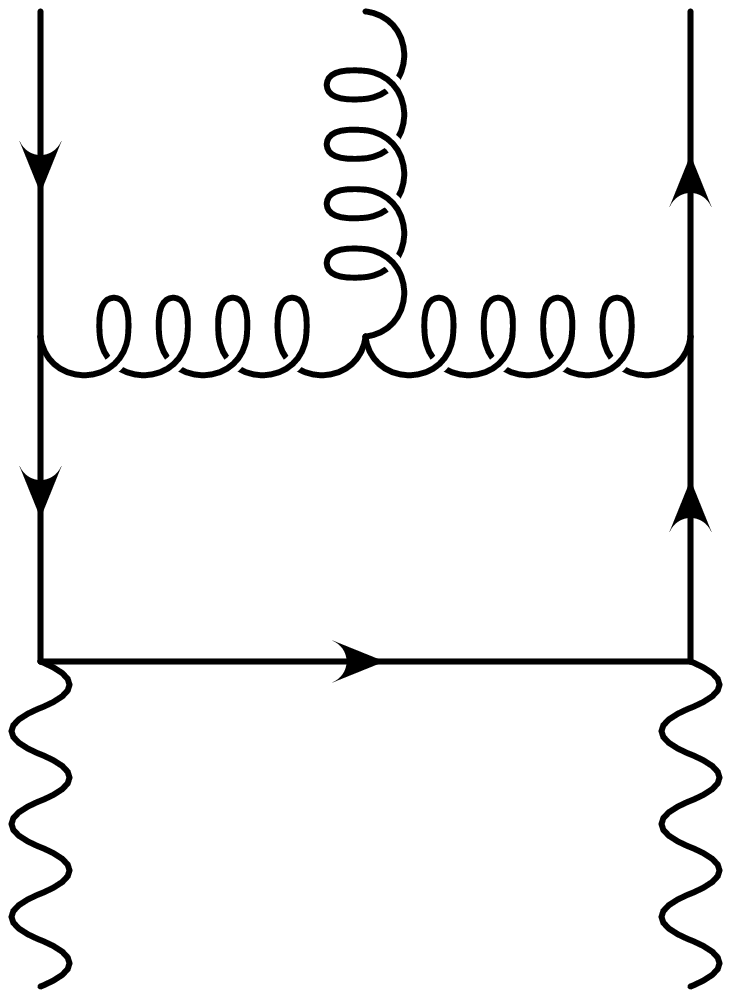} }
\resizebox{1.3cm}{2cm}{\includegraphics{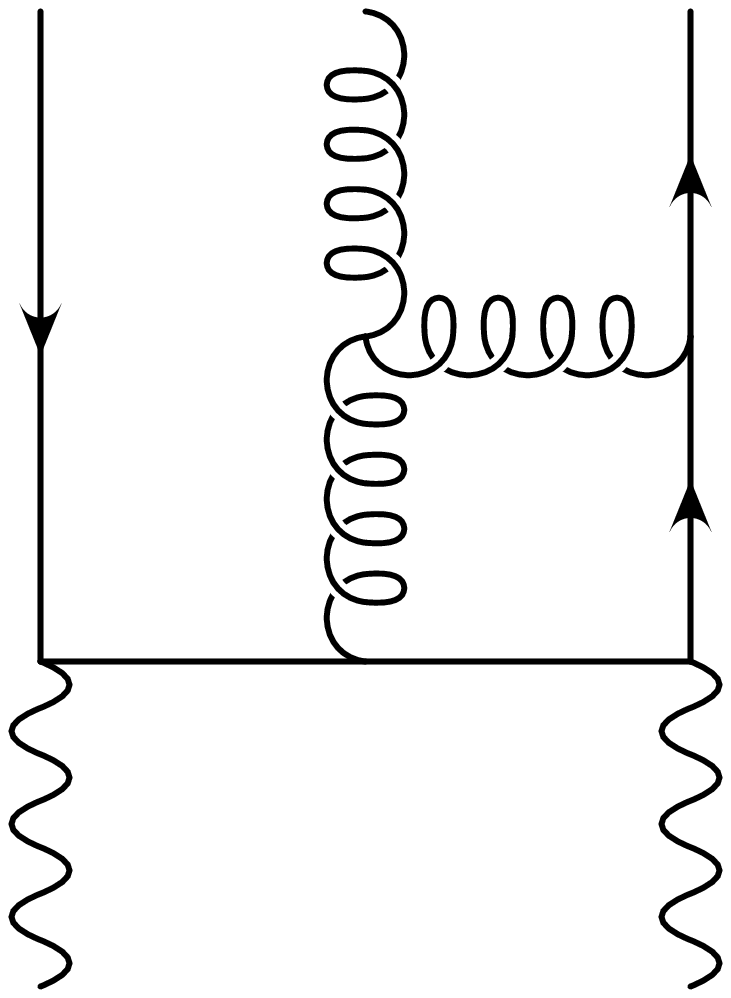} }
\resizebox{1.3cm}{2cm}{\includegraphics{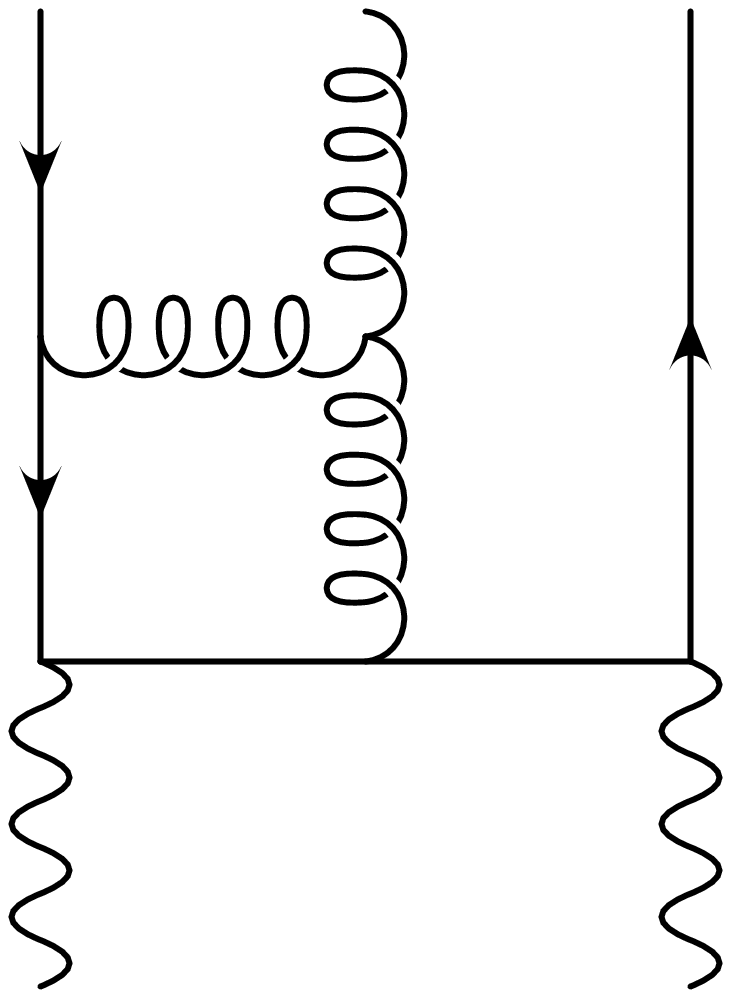} }
\resizebox{1.3cm}{2cm}{\includegraphics{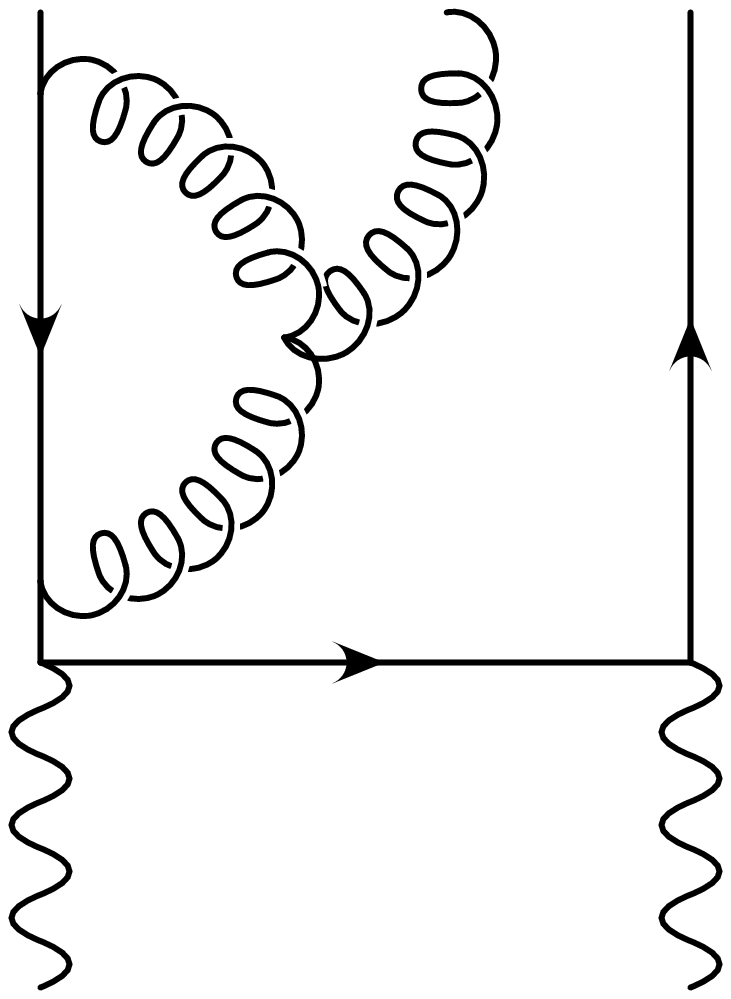} }
\resizebox{1.3cm}{2cm}{\includegraphics{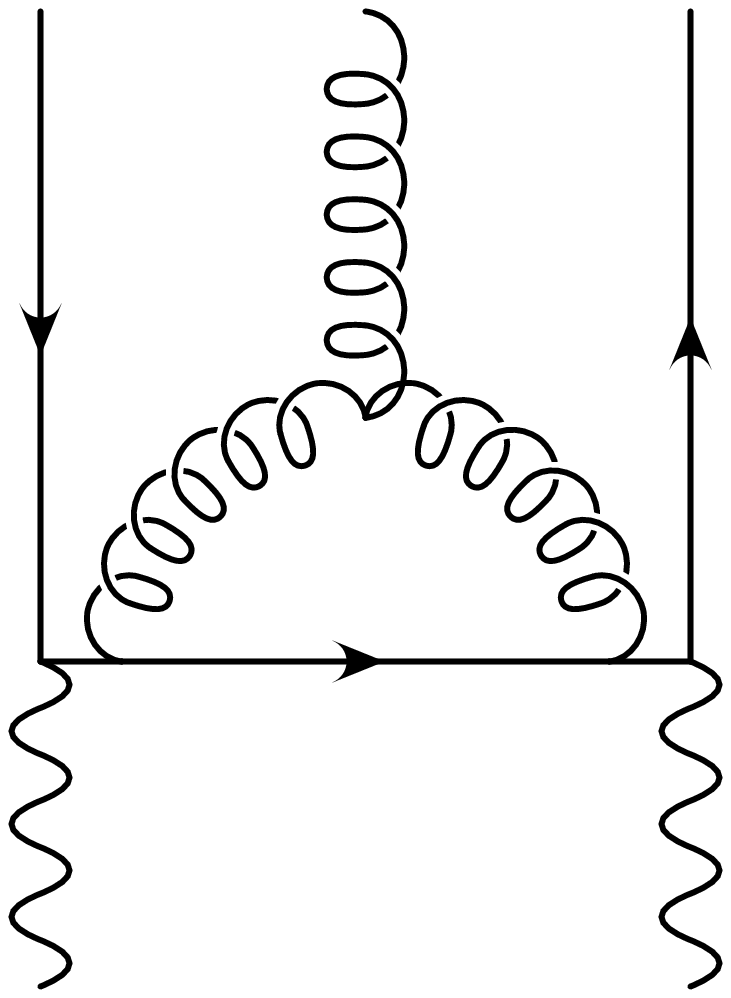} }
\resizebox{1.3cm}{2cm}{\includegraphics{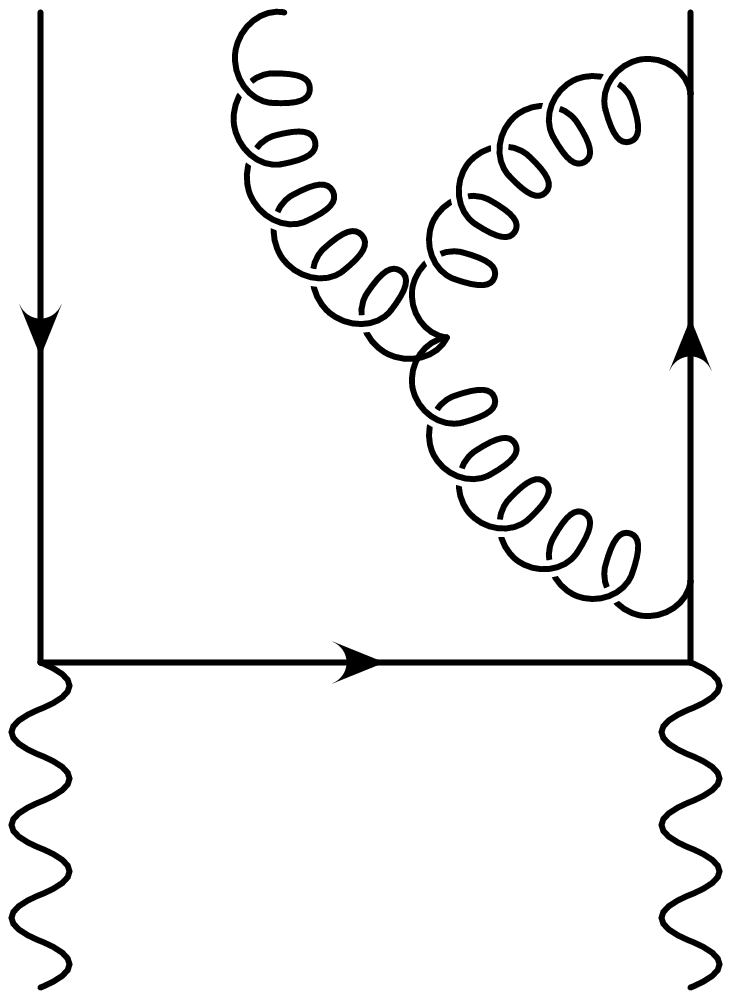} }
\resizebox{1.3cm}{2cm}{\includegraphics{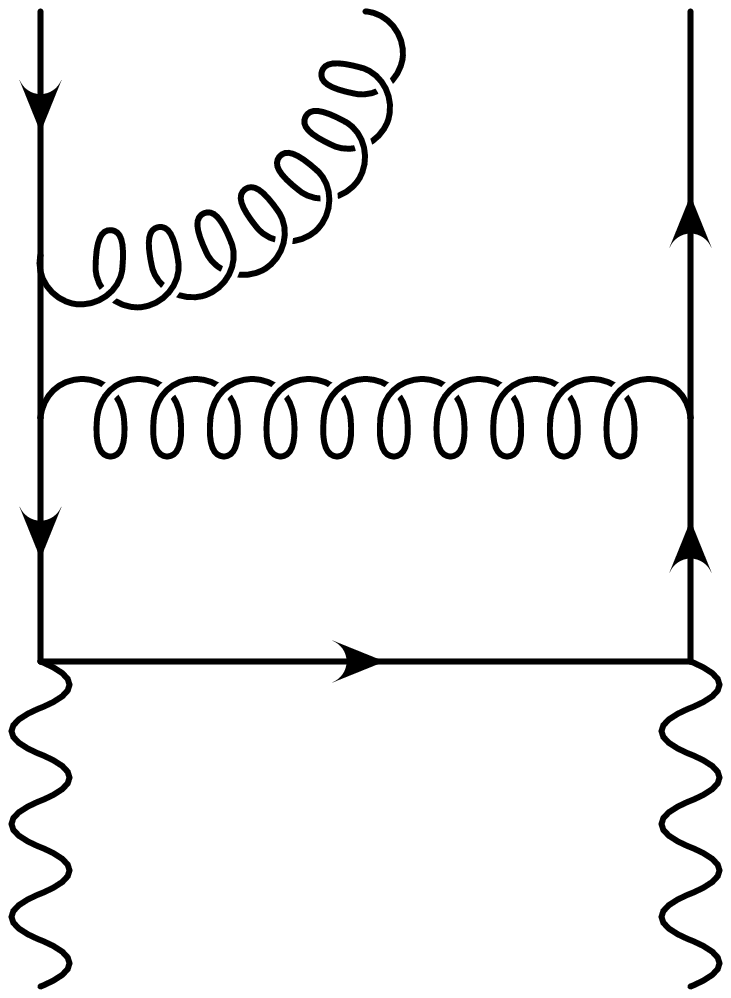} }
\resizebox{1.3cm}{2cm}{\includegraphics{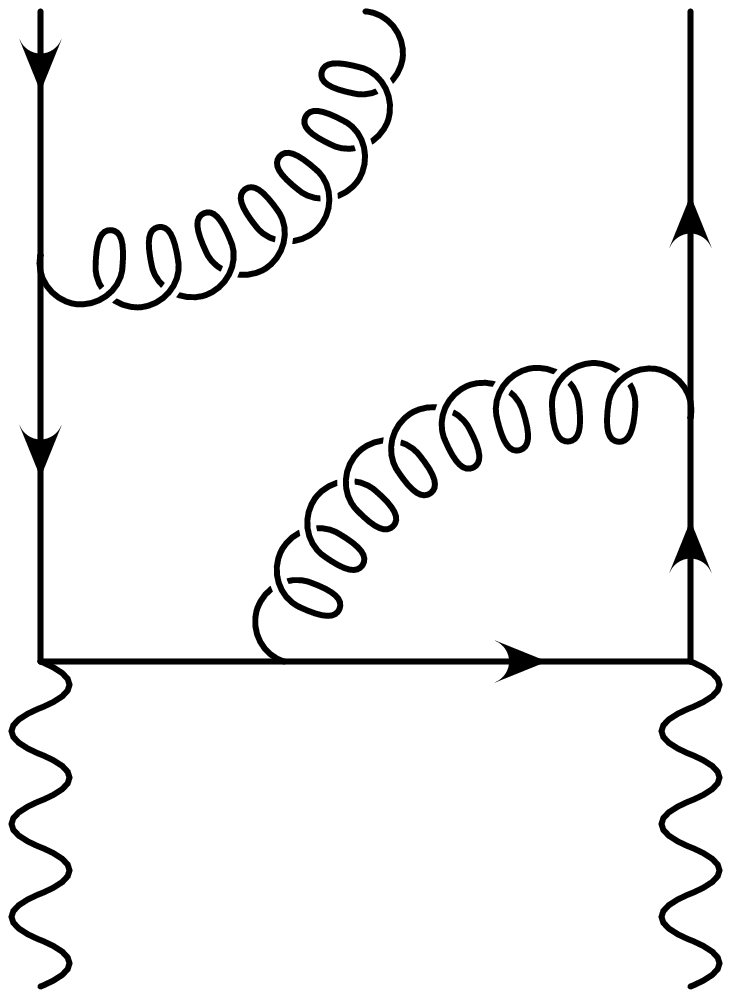} }
\resizebox{1.3cm}{2cm}{\includegraphics{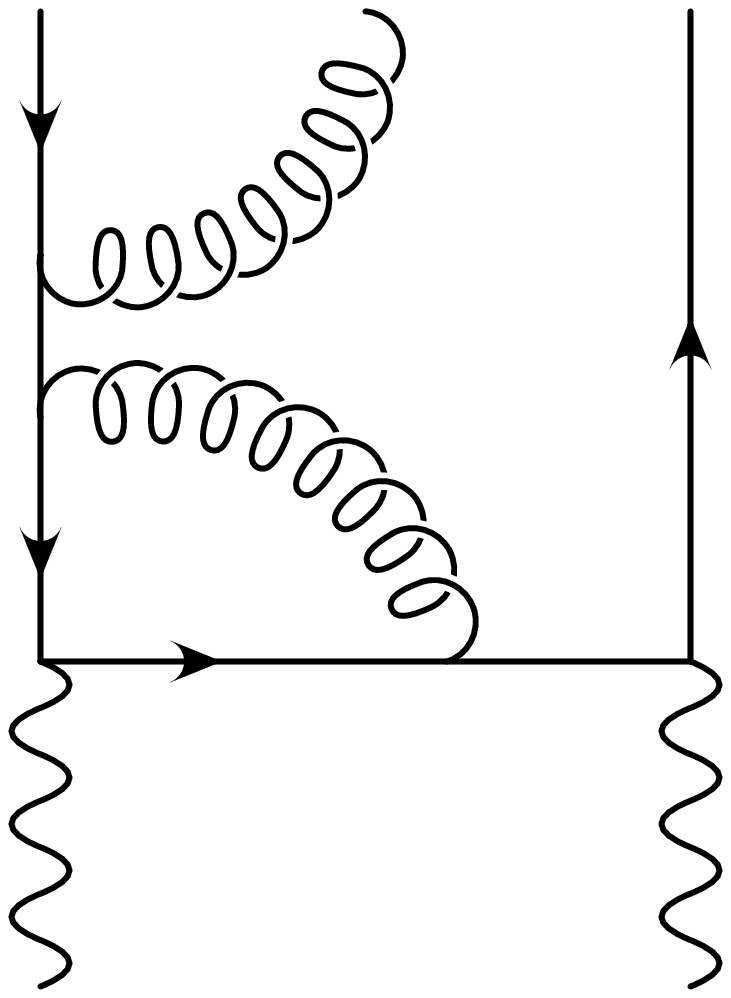} }
\resizebox{1.3cm}{2cm}{\includegraphics{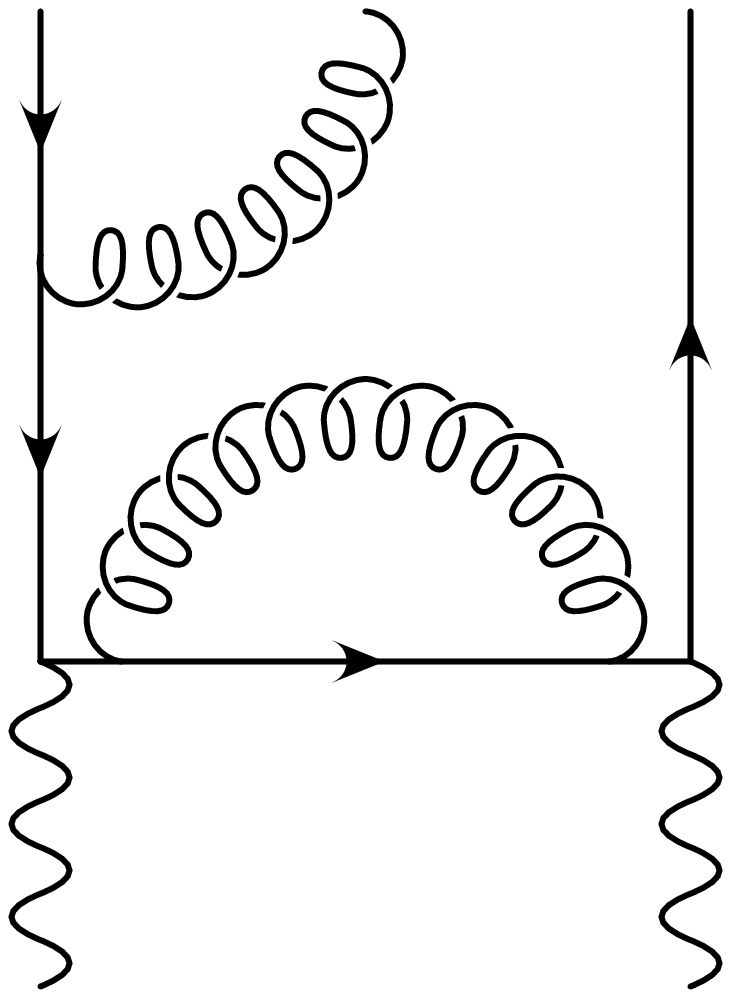} }

\resizebox{1.3cm}{2cm}{\includegraphics{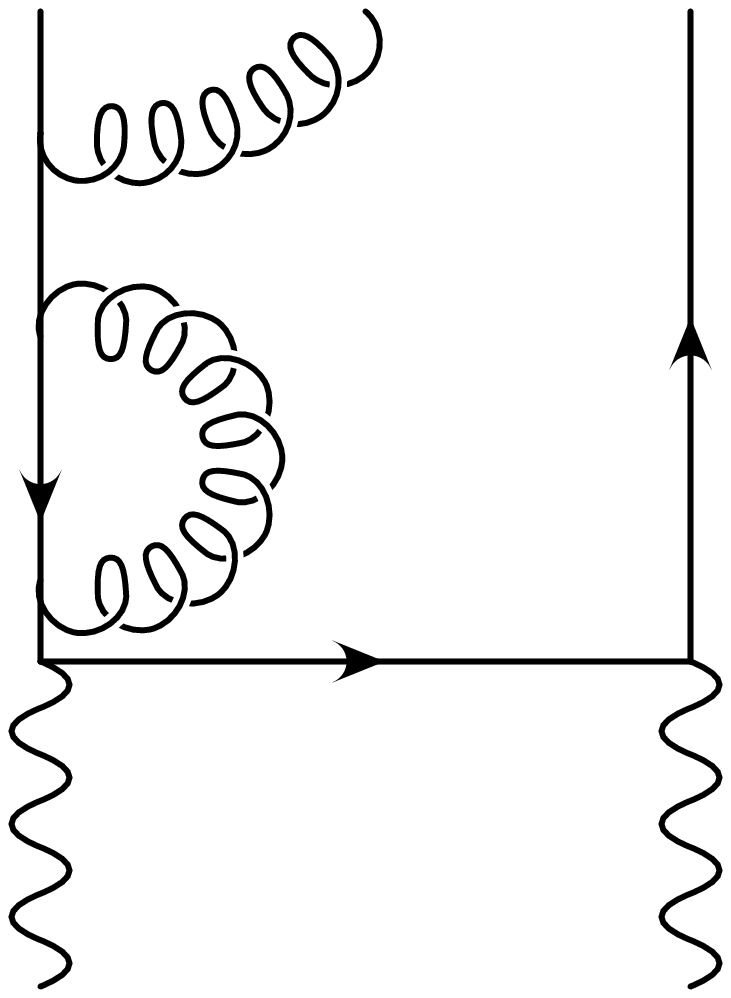} }
\resizebox{1.3cm}{2cm}{\includegraphics{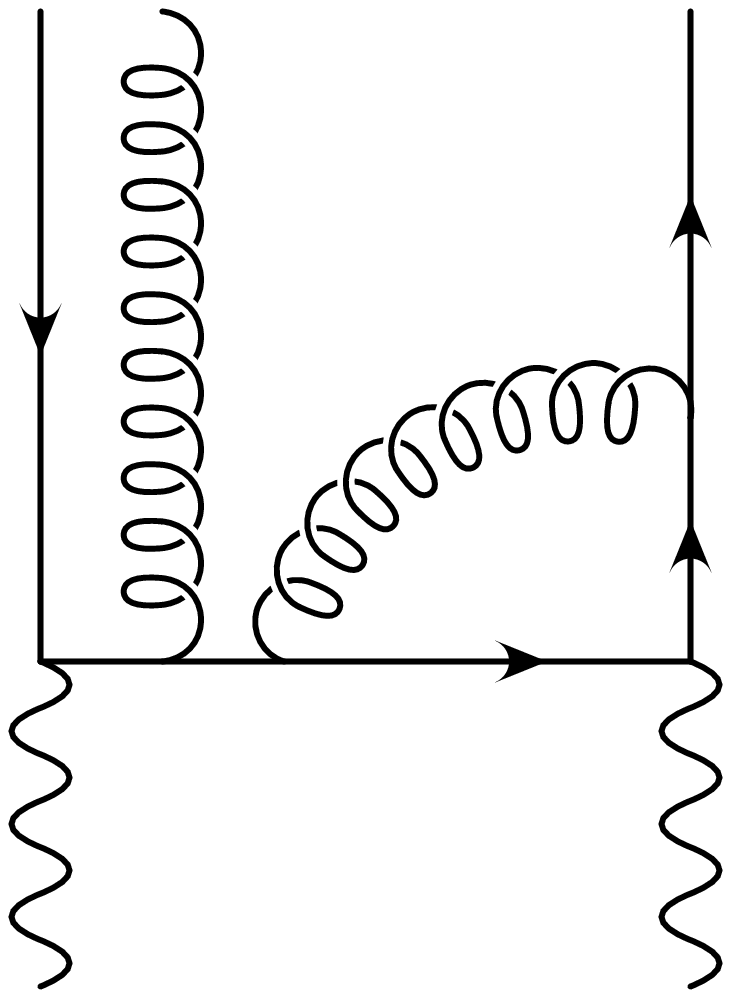} }
\resizebox{1.3cm}{2cm}{\includegraphics{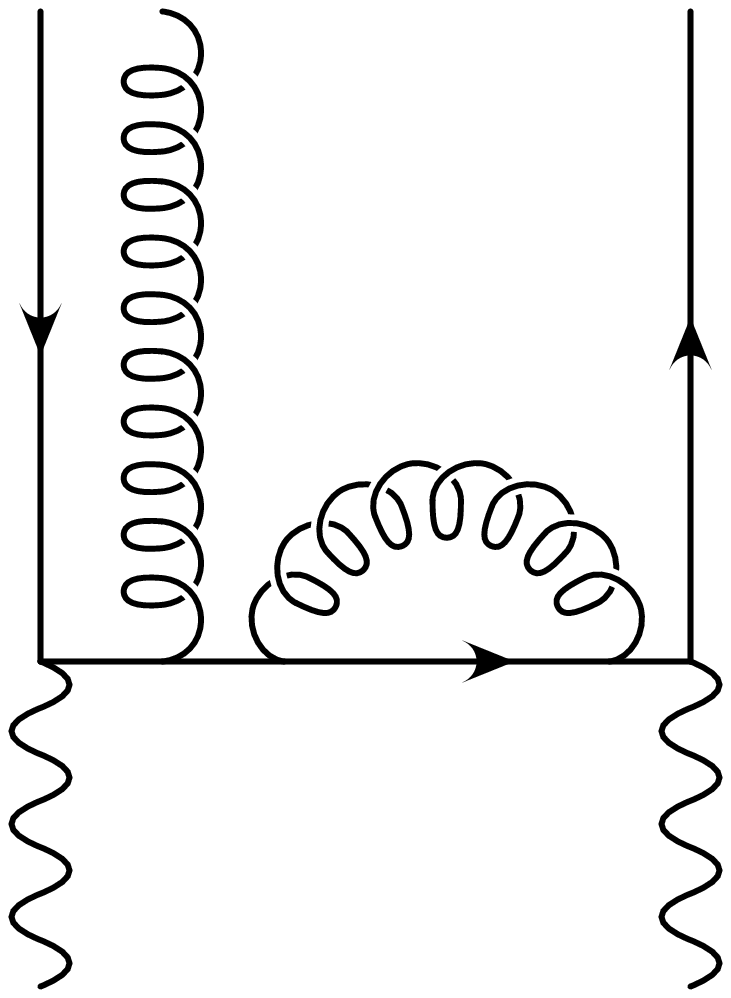} }
\resizebox{1.3cm}{2cm}{\includegraphics{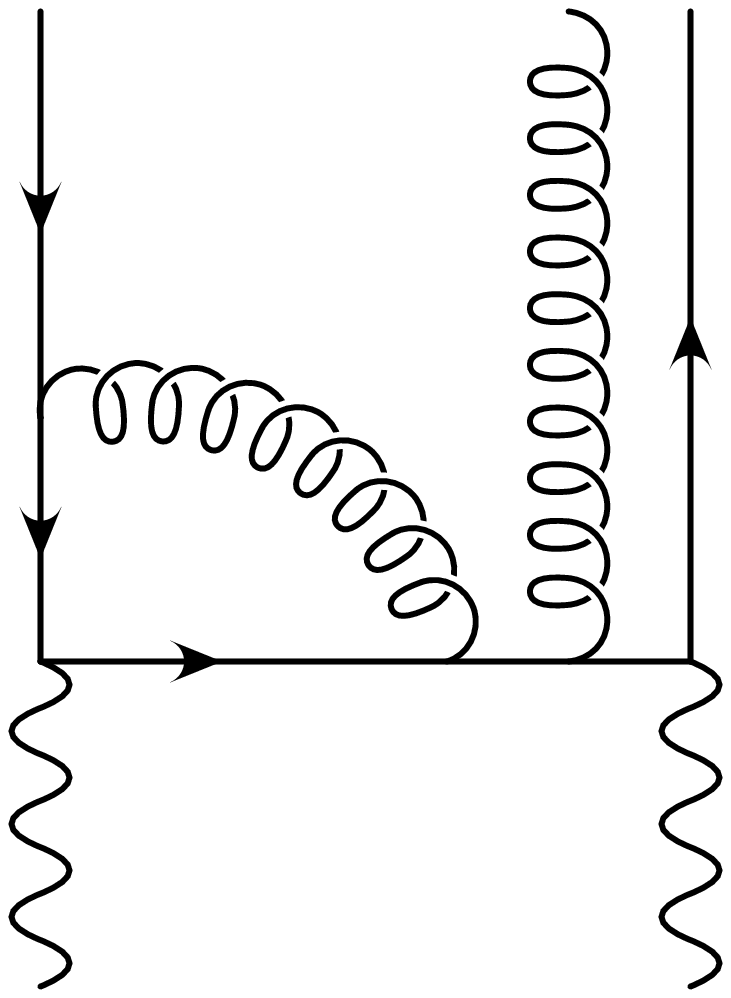} }
\resizebox{1.3cm}{2cm}{\includegraphics{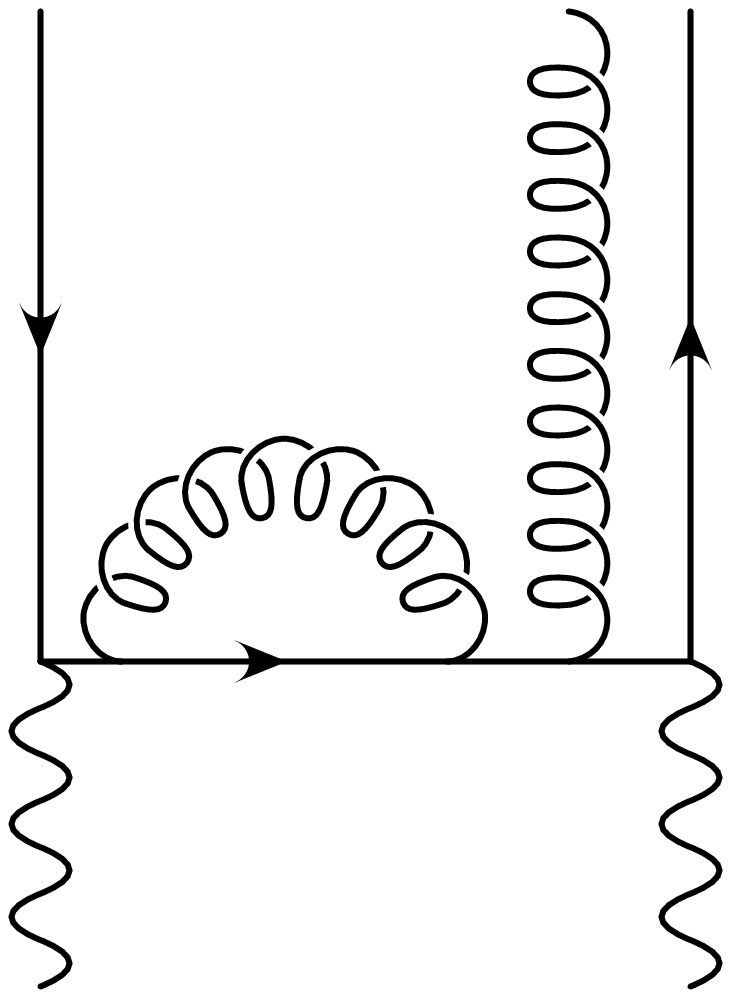} }
\resizebox{1.3cm}{2cm}{\includegraphics{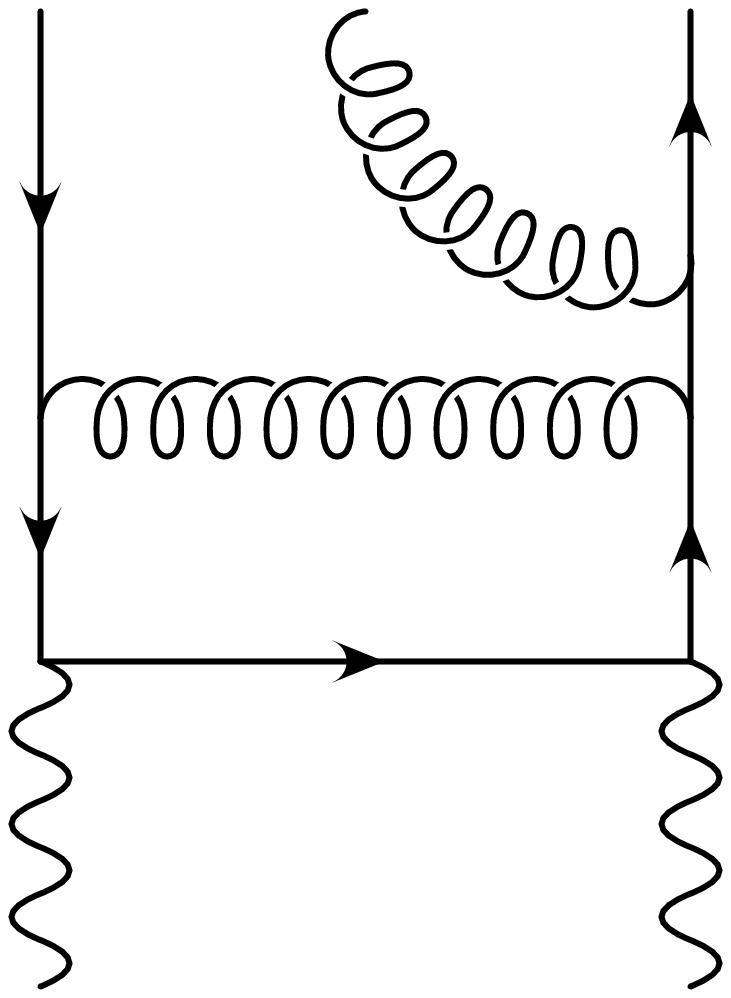} }
\resizebox{1.3cm}{2cm}{\includegraphics{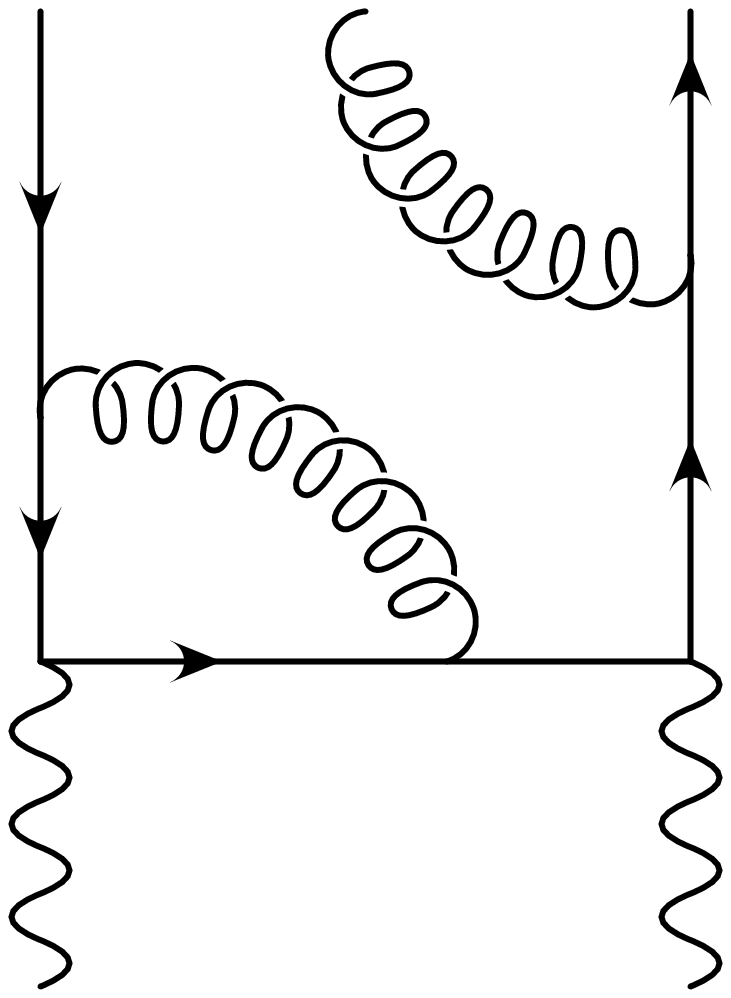} }
\resizebox{1.3cm}{2cm}{\includegraphics{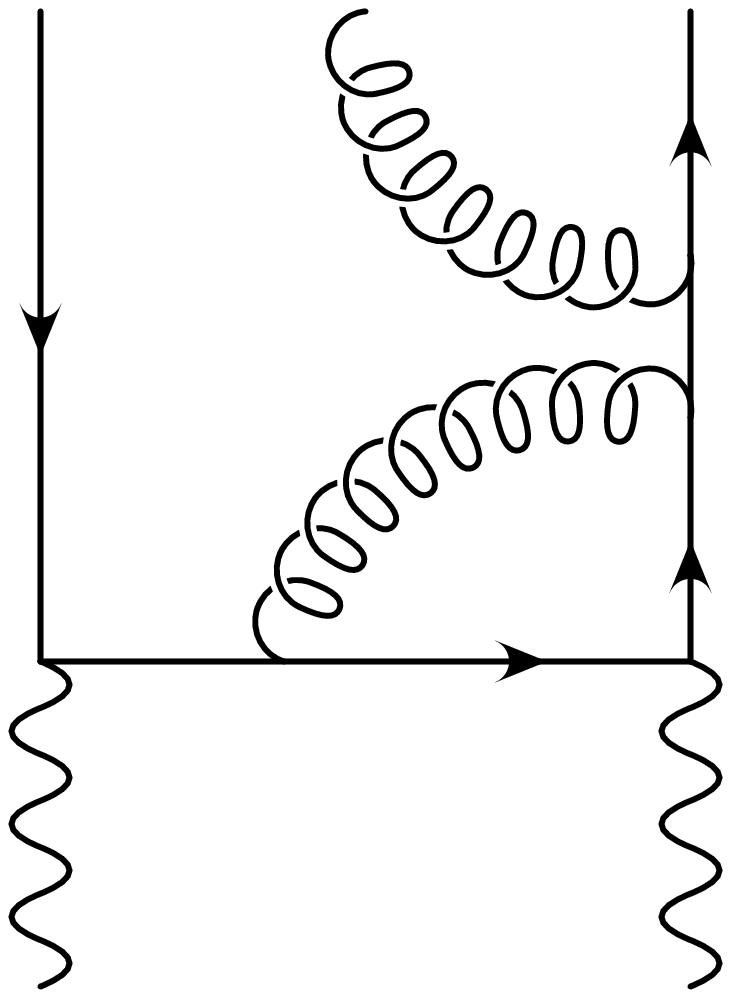} }
\resizebox{1.3cm}{2cm}{\includegraphics{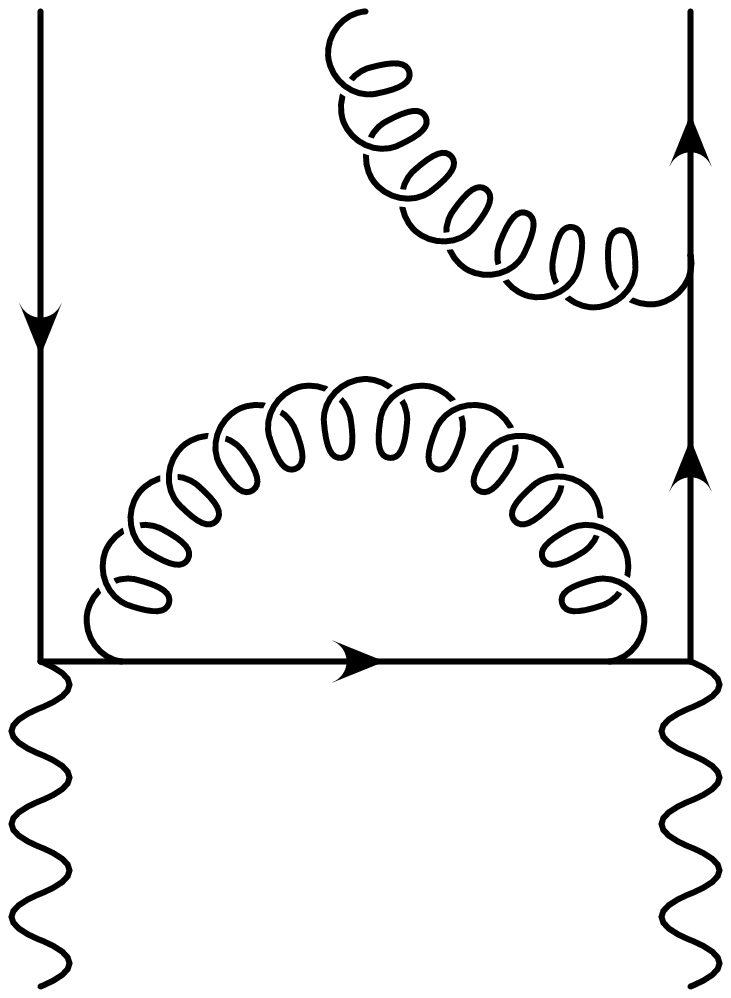} }
\resizebox{1.3cm}{2cm}{\includegraphics{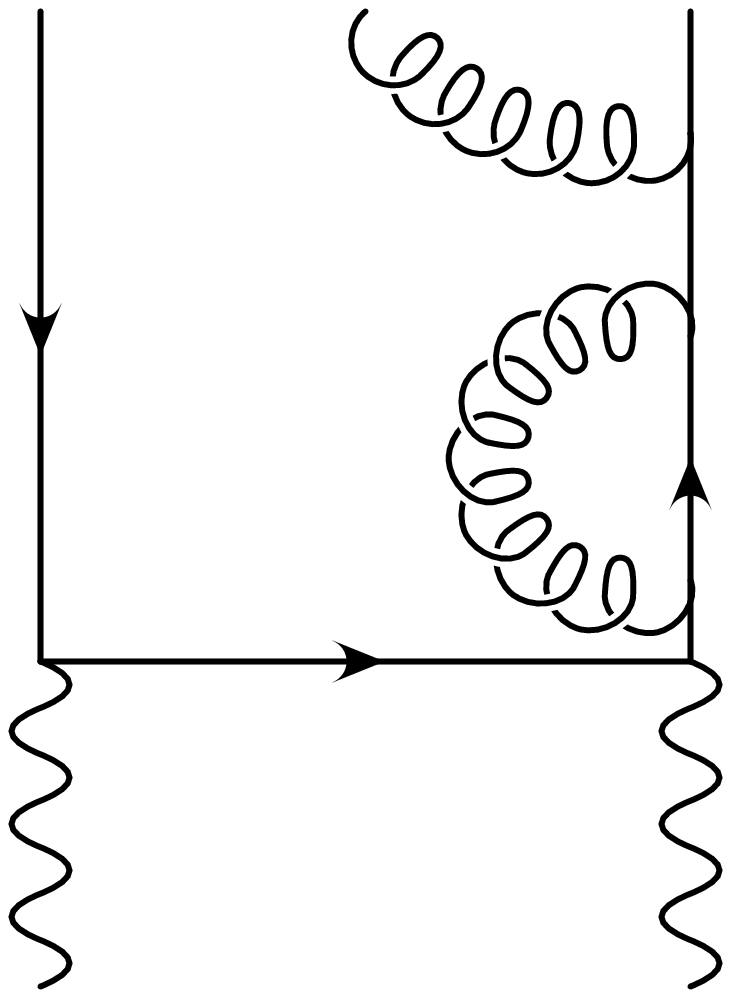} }

\caption{Feynman diagrams for ${\cal O}(N_c)$}
\label{NAfig}
\end{center}
\end{figure}

The Feynman diagrams which contribute to the 
${\cal O}(N_c)$-part are given in figure \ref{NAfig}. 
The amplitude of all helicity positive gluon and photons 
vanishes again at the tree level. 
The corresponding amplitude at one-loop level is 
infrared and ultraviolet finite as,
\bea
\lefteqn{
{\cal M}_5^{2}
(g_1^+\bar{q}_2^- \gamma_3^+\gamma_4^+ {q}_5^+)}
\nonumber\\
&=&
{-i\over (4\pi)^2}
{\sqrt{2}
\over 
[12]
\NB{51}\NB{14}
\NB{13}\NB{34}
}
\nonumber\\
&\times&\biggl[
{\frac {\PB{{23}}\left (-\NB{{12}}\NB{{34}}\PB{{14}}+\PB{{45}}\NB{{35}}\NB{{
24}}\right )s_{{12}}s_{{14}}}{s_{{45}}s_{{23}}}}
%\nonumber\\
%&~&
-{\frac {\left (\PB{{23}}\NB{{34}}\PB{{45}}\NB{{25}}+\PB{{12}}\NB{{23}}\PB{{
34}}\NB{{14}}\right )s_{{23}}s_{{34}}}{s_{{45}}s_{{12}}}}
\nonumber\\
&~&
-{\frac {\left (-\PB{{45}}\NB{{51}}\PB{{12}}\NB{{24}}+\NB{{23}}\PB{{34}}\NB{{
45}}\PB{{25}}+2\,\PB{{12}}\NB{{23}}\PB{{34}}\NB{{14}}\right )\left (s_{{51}
}-s_{{34}}\right )}{s_{{12}}}}
\nonumber\\
&~&
+{\frac {\left (\PB{{45}}\NB{{51}}\PB{{12}}\NB{{24}}+\NB{{12}}\PB{{23}}\NB{{34
}}\PB{{14}}-s_{{34}}s_{{12}}-2\,s_{{34}}s_{{23}}\right )\left (s_{{12}}
+s_{{51}}+s_{{34}}\right )}{s_{{45}}}}
\nonumber\\
&~&
+\NB{{45}}\PB{{14}}\NB{{12}}\PB{{25}}+\PB{{45}}\NB{{14}}\PB{{13}}\NB{{35}}-{s_{
{23}}}^{2}-s_{{34}}s_{{23}}+s_{{45}}s_{{23}}
-4\,s_{{23}}\left (s_{{45}}+s_{{34}}\right )
\biggr]
\nonumber\\
&~&
+ (3 \leftrightarrow 4).
\nonumber
\eea

Tree level amplitudes with the helicity configuration 
$(\bar{q},q,\gamma^+,\gamma^\pm,g^\mp)$ 
are non-vanishing. 
Corresponding one-loop amplitudes have the ultraviolet 
and infrared divergences, again.

The amplitude in which a gluon has negative helicity and 
two photons have positive helicity is,
\bea
\lefteqn{
{\cal M}_5^{2}
(g_1^- \bar{q}_2^-\gamma_3^+\gamma_4^+{q}_5^+)}
\nonumber\\
&=&
%{4\over i(4\pi)^2}
-c_\Gamma {\cal M}^{\mbox{tree}} 
\left\{
-{2\over \epsilon^2} 
\left({\mu^2\over -s_{12}}\right)^\epsilon
-{2\over \epsilon^2} 
\left({\mu^2\over -s_{51}}\right)^\epsilon
-{3\over \epsilon(1-2\epsilon)}
\left({\mu^2\over -s_{51}}\right)^\epsilon
-\delta_D 
\right\}
\nonumber\\
&-&{i\over (4\pi)^2}
{\sqrt{2}
\over 
[12]^2
\NB{34}^2\NB{25}
}
\Biggl[
2\,{\frac {\NB{{14}}{\PB{{12}}}^{2}\NB{{25}}s_{{34}}
\left (2\,\PB{{45}}\NB{{51}}\NB{
{24}}+3\,\NB{{23}}\PB{{34}}\NB{{14}}\right )
}{\NB{{45}}}}
Rln^2 \left( {s_{51}\over s_{23}}\right)
\nonumber\\
&~&
-4\,{\frac {\NB{{34}}\NB{{25}}{s_{{12}}}^{2}}{\NB{{23}}\NB{{45}}}}
(L(s_{51},s_{45},s_{23})+L(s_{23},s_{12},s_{45}))
+
4\,{\frac {\NB{{25}}\PB{{51}}\NB{{34}}\NB{{12}}s_{{12}}}{
\NB{{23}}\NB{{24}}}}
L(s_{12},s_{51},s_{34})
\nonumber\\
&~&
-2{
\NB{25}\PB{12}\PB{45}
(\NB{14}s_{51}s_{34}-\NB{24}\PB{25}\NB{51}s_{45})
s_{34}\over 
s_{51}s_{45}(s_{23}-s_{51})
}
~+~(3\leftrightarrow 4)
~\Biggr]
,
\nonumber
\eea
where
\[
{\cal M}^{\mbox{tree}} 
={\frac {
2\sqrt{2}
{i\NB{{12}}}^{2}\NB{{25}}}{\NB{{45}}\NB{{24}}\NB{{23}}\NB{{35}}}}.
\]
The tree level amplitudes are given by,
\[
 {\cal A}_5^{tree}(q\bar{q}g\gamma\gamma)
=e^2gT^a{\cal M}_5^{tree}(q\bar{q}g\gamma\gamma).
\]

The amplitude with 
the helicity $(\bar{q}^-,q^+,\gamma^+,\gamma^-,g^+)$ is 
given by, 
\bea
\lefteqn{
{\cal M}_5^{2}
(g_1^+\bar{q}_2^- \gamma_3^-\gamma_4^+{q}_5^+  )}
\nonumber\\
&=&
%{4\over i(4\pi)^2}
-c_\Gamma {\cal M}^{\mbox{tree}} 
\Biggl\{
-{2\over \epsilon^2} 
\left({\mu^2\over -s_{12}}\right)^\epsilon
-{2\over \epsilon^2} 
\left({\mu^2\over -s_{51}}\right)^\epsilon
-{3\over \epsilon(1-2\epsilon)}
\left({\mu^2\over -s_{51}}\right)^\epsilon
\nonumber\\
&&
-2 L(s_{34},s_{23},s_{51})-2 L(s_{12},s_{51},s_{34})
-2-\delta_D
\Biggr\}
\nonumber\\
&-&{i\over (4\pi)^2}
{\sqrt{2}
\over 
[23]^2
\NB{14}^2\NB{25}
}
\Biggl[
f_1 Rln^2\left({s_{35}\over s_{12}}\right) 
~+~f_2 Rln^2\left({s_{35}\over s_{24}}\right) 
~+~f_3 Rln^1\left({s_{34}\over s_{12}}\right) 
\nonumber\\
&&
+\Bigl\{
-4\,{\frac {\PB{{45}}\NB{{51}}\PB{{12}}\NB{{24}}\left (\PB{{45}}\NB{{51}}\PB{
{12}}\NB{{24}}-\NB{{23}}\PB{{34}}\NB{{45}}\PB{{25}}\right )^{2}}{s_{{12}}{s
_{{45}}}^{3}}}
\nonumber\\
&&
-2\,{\frac {\left (2\,\PB{{45}}\NB{{51}}\PB{{12}}\NB{{24}}-\PB{{51}}\NB{{25}
}\PB{{24}}\NB{{14}}-\NB{{23}}\PB{{34}}\NB{{45}}\PB{{25}}\right )
}{{s_{{45}}}^{2}}}
\nonumber\\
&&~~~~~~~~~~~~~~~~~~~~~~\times
\left (
2~\NB{{23}}\PB{{34}}\NB{{45}}\PB{{25}}+2~\PB{{23}}\NB{{34}}\PB{{45}}\NB{{25}}
+s_{12}s_{45}
\right )
\nonumber\\
&&
+2\,{\frac {\PB{{45}}\left (\NB{{24}}\PB{{12}}\NB{{51}}-\PB{{23}}\NB{{34}}\NB{
{25}}\right )s_{{51}}s_{{24}}}{{s_{{45}}}^{2}}}
%\nonumber\\
%&&
+2\,{\frac {\left (-s_{{51}}s_{{23}}-2\,s_{{34}}s_{{23}}+{s_{{51}}}^{2}
-s_{{34}}s_{{51}}\right )s_{{25}}}{s_{{45}}}}
\nonumber\\
&&
+2\,{\frac {\left (\NB{{45}}\PB{{51}}\NB{{12}}+\NB{{23}}\NB{{14}}\PB{{13}}
\right ){s_{{23}}}^{2}}{s_{{51}}\NB{{24}}}}
%\nonumber\\
%&&
-2\,{\frac {\PB{{51}}\left (\NB{{12}}\PB{{23}}\NB{{35}}+\NB{{13}}\NB{{45}}\PB
{{34}}\right )\left (-s_{{23}}+s_{{34}}\right )}{s_{{51}}}}
\nonumber\\
&&
+2\,{\frac {{s_{{23}}}^{2}\left (-s_{{23}}+s_{{45}}\right )}{s_{{51}}}}
%\nonumber\\
%&&
-2\,\PB{{24}}\left (\PB{{51}}\NB{{25}}\NB{{14}}-\NB{{34}}\NB{{12}}\PB{{13}}
\right )
+2\,{s_{{51}}}^{2}-6\,s_{{34}}s_{{51}}+2\,s_{{51}}s_{{23}}
\nonumber\\
&&
-2\,s_{{34}}s_{{45}}-4\,s_{{51}}s_{{12}}
+2\,s_{{45}}s_{{23}}+4\,{s_{{34}}}^{2}+2\,s
_{{12}}s_{{34}}\Bigr\}\tilde{L}(s_{34},s_{35},s_{12})
\nonumber\\
&&
~+~4\,{\frac {\PB{{14}}\PB{{25}}{s_{{23}}}^{2}L(s_{{51}},s_{{45}},s_{{23}})
}{\PB{{12}}\PB{{45}}}}
\nonumber\\
&&
+\Bigl\{
2\,{\frac {\PB{{51}}\left (\NB{{12}}\PB{{23}}\NB{{35}}+\PB{{34}}\NB{{45}}\NB{
{13}}\right ){s_{{34}}}^{2}\left (s_{{23}}-s_{{45}}\right )}{s_{{51}}s
_{{24}}s_{{14}}}}
%\nonumber\\
%&&
-4\,{\frac {\NB{{12}}\PB{{13}}{\NB{{34}}}^{2}\PB{{23}}\PB{{45}}\NB{{25}}}{\NB
{{24}}s_{{14}}}}
\nonumber\\
&&
+2\,{\frac {s_{{34}}\left (\NB{{12}}\PB{{13}}\PB{{24}}-\PB{{23}}\PB{{45}}\NB{
{25}}\right )\left (\PB{{12}}\NB{{23}}\NB{{14}}+\NB{{45}}\PB{{51}}\NB{{13}}
\right )}{s_{{24}}s_{{14}}}}
\nonumber\\
&&
-2\,{\frac {{\PB{{34}}}^{2}\left (\PB{{12}}\NB{{23}}\NB{{14}}+\NB{{45}}\PB{{
51}}\NB{{13}}\right )^{2}}{s_{{24}}s_{{14}}}}
\Bigr\}
\tilde{L}(s_{24},s_{12},s_{35})~+R_0 ~\Biggr],
\nonumber
\eea
where 
\[
 {\cal M}^{tree}=
{i2\sqrt{2}\NB{25}\NB{23}^2\over \NB{12}\NB{45}\NB{24}\NB{51}}.
\]
We also used following functions,
\bea
\tilde{L}(s_{34},s_{35},s_{12})
&=&{L}(s_{34},s_{35},s_{12})
-s_{45}Rln^1\left({s_{35}\over s_{12}}\right)
+{1\over2}s_{45}^2Rln^2\left({s_{35}\over s_{12}}\right)
-s_{45}Rln^1\left({s_{34}\over s_{12}}\right)
\nonumber\\
\tilde{L}(s_{24},s_{12},s_{35})
&=&{L}(s_{24},s_{12},s_{35})
+{1\over 2}s_{14}s_{24}Rln^2\left({s_{35}\over s_{12}}\right).
\nonumber
\eea
$f_i~(i=1,2,3)$ and $R_0$ are, 
\bea
f_1&=&
-\PB{{24}}\left (\NB{{45}}\PB{{51}}\NB{{12}}+\PB{{13}}\NB{{14}}\NB{{23}}
\right ){s_{{23}}}^{2}\left (-{\frac {-8\,s_{{45}}s_{{23}}+3\,{s_{{23}
}}^{2}+6\,{s_{{45}}}^{2}}{s_{{51}}s_{{24}}}}-{\frac {3\,s_{{51}}-6\,s_
{{23}}+8\,s_{{45}}}{s_{{24}}}}\right )
\nonumber\\
&&
-{\frac {\PB{{51}}\left (\NB{{13}}\NB{{45}}\PB{{34}}+\NB{{12}}\PB{{23}}\NB{{
35}}\right )\left (s_{{23}}-s_{{45}}\right )\left (3\,{s_{{23}}}^{2}-3
\,s_{{34}}s_{{23}}-5\,s_{{45}}s_{{23}}+3\,{s_{{34}}}^{2}+3\,s_{{34}}s_
{{45}}\right )}{s_{{51}}}}
\nonumber\\
&&
+{\frac {\PB{{12}}\left (\PB{{45}}\NB{{51}}\NB{{24}}-\NB{{14}}\PB{{34}}\NB{{23
}}\right ){s_{{23}}}^{2}s_{{45}}}{s_{{12}}}}
%\nonumber\\
%&&
+{\frac {{s_{{23}}}^{2}\left (s_{{23}}-s_{{45}}\right )\left (-8\,s_{{
45}}s_{{23}}+3\,{s_{{23}}}^{2}+6\,{s_{{45}}}^{2}\right )}{s_{{51}}}}
\nonumber\\
&&
+\left (2\,\NB{{23}}\PB{{34}}\NB{{45}}\PB{{25}}+2\,\NB{{23}}\PB{{24}}\NB{{14}}
\PB{{13}}-2\,\PB{{51}}\NB{{25}}\PB{{23}}\NB{{13}}-{s_{{23}}}^{2}\right )s_{
{23}}\left (2\,s_{{51}}+4\,s_{{45}}-5\,s_{{23}}\right )
\nonumber\\
&&
-3\,(\PB{{34}}\NB{{45}}\PB{{51}}\NB{{13}}+\NB{{23}}\PB{{24}}\NB{{14}}\PB
{{13}}-2\,\NB{{12}}\PB{{23}}\NB{{34}}\PB{{14}}
\nonumber\\
&&~~~~~~~~~~~~~~~~~~~~~~
+2\,s_{{34}}s_{{45}}+s_{{51}}s_{{23}}-2\,{s_{{23}}}^{2})
\left (-s_{{45}}+s_{{12}}\right )
\left (s_{{45}}+s_{{34}}\right )
\nonumber\\
&&
-s_{{23}} ({s_{{23}}}^{3}-6\,s_{{34}}{s_{{23}}}^{2}-{s_{{23}}}^{2
}s_{{12}}-s_{{51}}{s_{{23}}}^{2}+2\,{s_{{23}}}^{2}s_{{45}}-2\,s_{{51}}
s_{{23}}s_{{12}}-7\,s_{{23}}{s_{{45}}}^{2}
\nonumber\\
&&
-2\,s_{{23}}s_{{34}}s_{{45}}
+3\,s_{{23}}s_{{45}}s_{{12}}+6\,{s_{{34}}}^{2}s_{{45}}-2\,s_{{12}}{s_{
{45}}}^{2}+2\,{s_{{45}}}^{3}+6\,s_{{34}}{s_{{45}}}^{2})
\nonumber
\eea

\bea
f_2&=&
\PB{{51}}\left (\NB{{13}}\NB{{45}}\PB{{34}}+\NB{{12}}\PB{{23}}\NB{{35}}
\right )s_{{14}}
\nonumber\\
&&\times
\left (-{\frac {s_{{23}}\left (s_{{23}}-s_{{45}}
\right )}{s_{{51}}}}+3\,{\frac {s_{{34}}\left (s_{{23}}-s_{{45}}
\right )}{s_{{51}}}}+2\,{\frac {s_{{12}}\left (-s_{{23}}+s_{{34}}
\right )}{s_{{51}}}}\right )
\nonumber\\
&&
+3\,\left (\PB{{23}}\NB{{34}}\PB{{45}}\NB{{25}}-\PB{{51}}\NB{{12}}\PB{{23}}\NB{
{35}}+{s_{{23}}}^{2}\right )\left (s_{{12}}-s_{{45}}\right )s_{{14}}
\nonumber\\
&&
+2\,\left (\PB{{23}}\NB{{34}}\PB{{45}}\NB{{25}}-\PB{{51}}\NB{{12}}\PB{{23}}\NB{
{35}}-s_{{34}}s_{{12}}\right )\left (s_{{12}}+s_{{23}}\right )s_{{14}}
\nonumber\\
&&
+2\,\PB{{14}}\left (\NB{{12}}\PB{{23}}\NB{{34}}+\NB{{51}}\PB{{25}}\NB{{24}}
\right )s_{{12}}s_{{25}}
%\nonumber\\
%&&
-2\,s_{{14}}{s_{{23}}}^{2}\left (-s_{{23}}+s_{{51}}+s_{{12}}+s_{{34}}
\right )
\nonumber\\
&&
-s_{{14}}\left (-2\,s_{{51}}+2\,s_{{34}}+3\,s_{{23}}\right ){s_{{12}}}
^{2}
%\nonumber\\
%&&
-s_{{14}}\left (3\,s_{{34}}-s_{{23}}\right ){s_{{45}}}^{2}
\nonumber\\
&&
+s_{{14}}s_{{34}}\left (-4\,s_{{51}}+2\,s_{{34}}+5\,s_{{45}}\right )s_{
{12}}
%\nonumber\\
%&&
+2\,s_{{14}}\left (3\,s_{{23}}s_{{34}}s_{{45}}+{s_{{51}}}^{2}s_{{12}}
\right )
\nonumber
\eea

%\newpage
%\input test.txt

\bea
f_3&=&
-{\frac {\PB{{45}}\left (\NB{{51}}\PB{{12}}\NB{{24}}-\PB{{23}}\NB{{34}}\NB{{
25}}\right )s_{{23}}\left (s_{{34}}s_{{23}}-s_{{12}}s_{{23}}-2\,s_{{51
}}s_{{12}}\right )}
{s_{{45}}s_{{12}}}}
\nonumber\\
&&
+{\frac {\left (\PB{{51}}\NB{{25}}\PB{{23}}\NB{{13}}+\NB{{23}}\PB{{34}}\NB{
{45}}\PB{{25}}\right ){s_{{23}}}
(3~s_{{25}}s_{25}+2~(s_{51}-s_{34})s_{45})
}{s_{{51}}s_{{24}}}}
\nonumber\\
&&
-{\frac {s_{{45}}{s_{{23}}}^{3}\left (3\,s_{{23}}+3\,s_{{12}}-2\,s_{{
45}}\right )}{s_{{51}}s_{{24}}}}
%\nonumber\\
%&&
-2\,{\frac {s_{{23}}\left (-s_{{23}}-s_{{51}}+s_{{34}}\right ){s_{{45}
}}^{2}}{s_{{51}}}}
\nonumber\\
&&
+3\,{\frac {s_{{45}}{s_{{23}}}^{2}\left (s_{{25}}-s_{{23}}\right )}{s_{
{51}}}}
%\nonumber\\
%&&
+3\,{\frac {\left (\NB{{23}}\PB{{34}}\NB{{45}}\PB{{25}}-\NB{{23}}\PB{{24}}\NB{
{14}}\PB{{13}}-{s_{{12}}}^{2}\right ){s_{{23}}}^{2}}{s_{{24}}}}
\nonumber\\
&&
-{\frac {{s_{{23}}}^{2}\left (6\,s_{{12}}s_{{23}}-2\,s_{{45}}s_{{12}}-
5\,s_{{45}}s_{{23}}+3\,{s_{{23}}}^{2}\right )}{s_{{24}}}}
%\nonumber\\
%&&
-{\frac {{s_{{23}}}^{2}s_{{34}}\left (-s_{{51}}+s_{{34}}\right )}{s_{{
12}}}}
\nonumber\\
&&
+4\,\left (-\NB{{23}}\PB{{24}}\NB{{14}}\PB{{13}}+2\,\PB{{51}}\NB{{25}}\PB{{23}
}\NB{{13}}+\PB{{12}}\NB{{23}}\PB{{34}}\NB{{14}}\right )s_{{23}}
\nonumber\
\eea

\bea
R_0&=&
\left (\NB{{51}}\PB{{12}}\NB{{23}}\PB{{35}}-\NB{{45}}\PB{{51}}\NB{{12}}\PB{{
24}}+s_{{12}}s_{{35}}-s_{{51}}s_{{35}}\right )\left ({\frac {s_{{12}}
\left (s_{{51}}+s_{{45}}\right )}{s_{{35}}\left (s_{{12}}+s_{{14}}
\right )}}-{\frac {s_{{12}}\left (-s_{{51}}+s_{{12}}\right )}{s_{{51}}
\left (s_{{12}}+s_{{14}}\right )}}\right )
\nonumber\\
&&
+\left (-\NB{{45}}\PB{{51}}\NB{{12}}\PB{{24}}-\NB{{23}}\NB{{14}}\PB{{13}}\PB{{
24}}+{s_{{14}}}^{2}\right )\left (-{\frac {s_{{14}}s_{{12}}\left (s_{{
14}}+s_{{45}}\right )}{s_{{45}}s_{{35}}\left (s_{{12}}-s_{{35}}\right 
)}}-{\frac {s_{{14}}\left (s_{{51}}+s_{{45}}\right )}{s_{{35}}\left (s
_{{12}}-s_{{35}}\right )}}\right )
\nonumber\\
&&
-{\frac {\left (\NB{{45}}\PB{{51}}\NB{{12}}\PB{{24}}+\PB{{12}}\NB{{23}}\PB{{
34}}\NB{{14}}\right )\left (\left (s_{{51}}-s_{{23}}\right )^{2}
-s_{{45}}s_{{14}}\right )}{s_{{45}}s_{{35}}}}
\nonumber\\
&&
+2\,{\frac {{s_{{14}}}^{2}s_{{45}}\left (s_{{51}}+s_{{45}}\right )}{s_{
{35}}\left (-s_{{12}}+s_{{35}}\right )}}-{\frac {{s_{{14}}}^{2}s_{{51}
}\left (s_{{45}}s_{{12}}-{s_{{45}}}^{2}-s_{{51}}s_{{45}}-s_{{14}}s_{{
12}}\right )}{s_{{45}}s_{{35}}\left (-s_{{12}}+s_{{35}}\right )}}
%\nonumber\\
%&&
-2\,{\frac {{s_{{51}}}^{3}s_{{12}}}{s_{{45}}s_{{35}}}}
\nonumber\\
&&
+2\,{\frac {\PB{{24}}\left (\NB{{45}}\PB{{51}}\NB{{12}}+\NB{{25}}\PB{{35}}\NB{
{34}}\right )\left (-s_{{34}}+s_{{51}}\right )^{2}}{s_{{51}}s_{{24}}}}
%\nonumber\\
%&&
-2\,{\frac {{s_{{12}}}^{2}\left (-s_{{45}}s_{{35}}+{s_{{51}}}^{2}+s_{{
51}}s_{{45}}\right )}{s_{{35}}\left (s_{{12}}+s_{{14}}\right )}}
\nonumber\\
&&
+{\frac {\left (s_{{45}}s_{{12}}-{s_{{23}}}^{2}\right )s_{{14}}}{s_{{35
}}}}
%\nonumber\\
%&&
+{\frac {\left (\PB{{45}}\NB{{51}}\PB{{12}}\NB{{24}}+\PB{{51}}\NB{{14}}\PB{{34
}}\NB{{35}}+{s_{{51}}}^{2}-{s_{{23}}}^{2}\right )s_{{51}}}{s_{{35}}}}
\nonumber\\
&&
-2\,{\frac {\left (\NB{{45}}\PB{{51}}\NB{{12}}\PB{{24}}+\PB{{12}}\NB{{23}}\PB
{{34}}\NB{{14}}\right ){s_{{23}}}^{2}}{s_{{45}}s_{{12}}}}
%\nonumber\\
%&&
+{\frac {s_{{23}}\left ({s_{{51}}}^{2}+{s_{{23}}}^{2}\right )}{s_{{45}}
}}
-2\,{\frac {s_{{45}}\left (s_{{12}}s_{{23}}-{s_{{34}}}^{2}\right )}{s_
{{51}}}}
\nonumber\\
&&
+{\frac {\left (\NB{{51}}\PB{{12}}\NB{{23}}\PB{{35}}+\NB{{12}}\PB{{13}}\NB{{35
}}\PB{{25}}+{s_{{45}}}^{2}\right )\left (-s_{{23}}+s_{{45}}+s_{{12}}
\right )}{s_{{51}}}}
-{\frac {{s_{{45}}}^{2}\left (s_{{23}}+s_{{34}}\right )}{s_{{
51}}}}
\nonumber\\
&&
-2\,{\frac {\left
(-\PB{{51}}\NB{{14}}\PB{{34}}\NB{{35}}+\NB{{34}}\PB{{35}}
\NB{{25}}\PB{{24}}-s_{{34}}s_{{45}}\right )\left (s_{{23}}-s_{{34}}
\right )}{s_{{51}}}}
-{\frac {-s_{{34}}{s_{{23}}}^{2}+{s_{{23}}}^{2}s_{{12}}}{s_{{51}
}}}
\nonumber\\
&&
+{\frac {{s_{{45}}}^{3}\left (s_{{13}}+s_{{24}}\right )}{s_{{51}}s_{{35
}}}}
%\nonumber\\
%&&
-7\,{\frac {s_{{34}}s_{{45}}\left (s_{{51}}+s_{{45}}\right )}{s_{{35}}
}}-6\,{\frac {s_{{23}}{s_{{45}}}^{2}+{s_{{51}}}^{2}s_{{34}}}{s_{{35}}}
}-4\,{\frac {s_{{51}}s_{{45}}\left (s_{{23}}+s_{{34}}\right )}{s_{{35}
}}}
\nonumber\\
&&
-2\,\PB{{12}}\NB{{23}}\PB{{34}}\NB{{14}}-2\,\NB{{34}}\PB{{35}}\NB{{25}}\PB{{24
}}-4\,s_{{34}}s_{{45}}+2\,s_{{14}}s_{{51}}-2\,{s_{{23}}}^{2}+4\,s_{{34
}}s_{{23}}-{s_{{45}}}^{2}
\nonumber\\
&&
-4\,{s_{{51}}}^{2}-2\,{s_{{34}}}^{2}-s_{{45}}
s_{{12}}-s_{{51}}s_{{12}}-s_{{12}}s_{{23}}-3\,s_{{51}}s_{{45}}-s_{{45}
}s_{{23}}+s_{{34}}s_{{12}}
\nonumber
\eea

\subsection{
${\cal M}_5^3(\bar{q}qg\gamma\gamma)$}

In the end of this section, we also give amplitudes 
which come from the fermion loop contributions. 
Here, we only consider massless quark loops. 

\bea
\lefteqn{
{\cal M}_5^{3}
( {q}_1^+ \bar{q}_2^- g_3^+\gamma_4^+\gamma_5^+)}
\nonumber\\
&=&
{16\over i\sqrt{2}(4\pi)^2}
%{
%\over 
%}
{\frac {\NB{{45}}\PB{{51}}\NB{{12}}\PB{{24}}s_{{14}}-\NB{{45}}\PB{{51}}\NB
{{12}}\PB{{24}}s_{{24}}-s_{{25}}s_{{24}}s_{{14}}+{s_{{24}}}^{2}s_{{51}}
}{s_{{12}}[24]
\NB{41}\NB{35}
\NB{54}\NB{43}}}
\nonumber
\eea

\bea
\lefteqn{
{\cal M}_5^{3}
({q}_1^+ \bar{q}_2^- g_3^+\gamma_4^-\gamma_5^+)
=
{\cal M}_5^{3}
({q}_1^+ \bar{q}_2^- \gamma_3^+g_4^-\gamma_5^+)
}
\nonumber\\
&=&
{1\over i(4\pi)^2}
{8\over \sqrt{2}\PB{24}^2\NB{35}^2\NB{12}}
\Biggl[
-2\,{s_{{24}}}^{2}\left (-s_{{12}}-s_{{35}}\right )s_{{35}}
Rln^2\left({s_{34}\over s_{12}}\right)
\nonumber\\
&&
+
\Biggl\{
2\,{\frac {\left (-\NB{{51}}\PB{{12}}\NB{{23}}\PB{{35}}-\NB{{45}}\PB{{14}}\NB
{{13}}\PB{{35}}+s_{{23}}s_{{12}}\right )s_{{14}}}{s_{{35}}}}
\nonumber\\
&&
+
2\,{\frac {\left (\NB{{12}}\PB{{23}}\NB{{34}}\PB{{14}}-\NB{{51}}\PB{{12}}\NB{
{23}}\PB{{35}}\right )s_{{12}}s_{{23}}}{\left (-s_{{12}}+s_{{45}}
\right )s_{{35}}}}
%\nonumber\\
%&&
+2\,{\frac {{s_{{34}}}^{2}\left (-{s_{{23}}}^{2}+s_{{12}}s_{{51}}+s_{{
23}}s_{{51}}\right )}{\left (-s_{{12}}+s_{{45}}\right )^{2}}}
\nonumber\\
&&
-{\frac {\left (\NB{{12}}\PB{{13}}\NB{{34}}\PB{{24}}-\NB{{51}}\PB{{12}}\NB{{
23}}\PB{{35}}-s_{{34}}s_{{23}}+2\,s_{{34}}s_{{51}}\right )\left (s_{{34
}}s_{{12}}+s_{{34}}s_{{23}}+s_{{23}}s_{{12}}\right )}{\left (-s_{{12}}
+s_{{45}}\right )^{2}}}
\nonumber\\
&&
-{\frac {\left (\NB{{23}}\PB{{34}}\NB{{45}}\PB{{25}}+\NB{{12}}\PB{{23}}\NB{{
34}}\PB{{14}}\right )\left (-s_{{34}}+s_{{51}}+2\,s_{{23}}\right )}{-s_
{{12}}+s_{{45}}}}
\nonumber\\
&&
-2\,{\frac {\PB{{13}}\left (\NB{{12}}\NB{{34}}\PB{{24}}-\NB{{23}}\PB{{25}}\NB
{{51}}\right )s_{{12}}}{-s_{{12}}+s_{{45}}}}
%\nonumber\\
%&&
-{\frac {s_{{23}}\left (-{s_{{34}}}^{2}+s_{{34}}s_{{51}}+2\,{s_{{12}}}
^{2}+4\,s_{{12}}s_{{51}}\right )}{-s_{{12}}+s_{{45}}}}
\nonumber\\
&&
+2\,\PB{{34}}\NB{{35}}\PB{{51}}\NB{{14}}+2\,\NB{{45}}\PB{{14}}\NB{{13}}\PB{{35}
}+2\,\PB{{45}}\NB{{14}}\PB{{12}}\NB{{25}}-2\,\NB{{23}}\PB{{24}}\NB{{45}}\PB{{
35}}
\nonumber\\
&&
-4\,s_{{23}}s_{{51}}+4\,s_{{34}}s_{{23}}-4\,s_{{23}}s_{{12}}+2\,s_{{51
}}s_{{45}}+{s_{{51}}}^{2}+{s_{{34}}}^{2}-2\,s_{{34}}s_{{12}}
\Biggr\}
%{ln\left({s_{14}\over s_{45}}\right)\over (s_{12}-s_{45})^2}
\nonumber\\
&&
~~~~~~~~~~~~~\times
\left\{ln\left({s_{45}\over s_{12}}\right)
-{(s_{45}-s_{12})^2\over (s_{34}-s_{12})^2}
ln\left({s_{34}\over s_{12}}\right)\right\}
\nonumber\\
&&
-2\,{\PB{{35}}}^{2}{\PB{{24}}}^{2}\left ({\NB{{34}}}^{2}{\NB{{25}}
}^{2}+{\NB{{23}}}^{2}{\NB{{45}}}^{2}\right )
\left\{
{
L(s_{{34}},s_{{45}},s_{{12}})
\over s_{35}^2}
-{(2s_{45}-3_{12}+2s_{34}) \over s_{35}}
Rln^2\left({34\over 12}\right)
\right\}
\nonumber\\
&&
{\frac {\left (\NB{{23}}\PB{{34}}\NB{{45}}\PB{{25}}+\NB{{12}}\PB{{23}}\NB{{34
}}\PB{{14}}\right )s_{{23}}s_{{34}}}{s_{{12}}\left (-s_{{12}}+s_{{45}}
\right )}}
%\nonumber\\
%&&
-{\frac {\left (\NB{{34}}\PB{{45}}\NB{{51}}\PB{{13}}+\NB{{45}}\PB{{51}}\NB{{
12}}\PB{{24}}\right )s_{{45}}s_{{51}}}{s_{{12}}\left (s_{{34}}-s_{{12}}
\right )}}
\nonumber\\
&&
{\frac {\left (\NB{{12}}\PB{{13}}\NB{{34}}\PB{{24}}-\NB{{51}}\PB{{12}}\NB{{23
}}\PB{{35}}\right )\left (s_{{34}}+s_{{23}}\right )}{-s_{{12}}+s_{{45}}
}}
%\nonumber\\
%&&
+2\,{\frac {s_{{23}}s_{{51}}s_{{34}}}{-s_{{12}}+s_{{45}}}}
\nonumber\\
&&
+{\frac {\left (-\NB{{45}}\PB{{51}}\NB{{12}}\PB{{24}}-\NB{{51}}\PB{{12}}\NB{{
23}}\PB{{35}}-s_{{12}}s_{{45}}\right )\left (-s_{{45}}+s_{{51}}\right )
}{s_{{34}}-s_{{12}}}}
\nonumber\\
&&
+2\,{\frac {s_{{51}}s_{{45}}\left (-s_{{45}}+s_{{23}}+s_{{12}}\right )}
{s_{{34}}-s_{{12}}}}
%\nonumber\\
%&&
+{\frac {\left (s_{{45}}+s_{{34}}\right )\left (-s_{{45}}+s_{{23}}
\right )\left (-s_{{34}}+s_{{51}}\right )}{s_{{12}}}}
\nonumber\\
&&
+{\frac {\left (\NB{{12}}\PB{{23}}\NB{{34}}\PB{{14}}-\NB{{45}}\PB{{51}}\NB{{12
}}\PB{{24}}+s_{{45}}s_{{34}}\right )\left (-s_{{45}}-s_{{34}}+s_{{23}}+
s_{{51}}\right )}{s_{{12}}}}
\nonumber\\
&&
+{s_{{34}}}^{2}-2\,s_{{34}}s_{{51}}+{s_{{45}}}^{2}+4\,s_{{34}}s_{{23}}+
{s_{{51}}}^{2}+{s_{{23}}}^{2}
\Biggr].
\nonumber
\eea

%\input chap34.tex
%\subsection{Cross check}

Now, we give some comments concerning the cross-check of 
our results.  
Besides the previous results, 
we also carried out the direct calculation of 
the amplitudes with the helicities 
$(\bar{q}^-, q^+, g^\mp, \gamma^-, \gamma^\pm)$ and 
$(\bar{q}^-, q^+, g^-, \gamma^-, \gamma^-)$. 
These amplitudes are obtained from the previous results 
by using Parity inversion and charge conjugation. 
This procedure is basically the same procedure 
discussed in section 2. 
Our direct calculations are consistent with this argument. 

In addition, we checked the singular parts. 
Singular parts of one-loop amplitudes have the 
well known universal structure eq.(\ref{eq:SIN})\cite{KUN,BC}. 
Thus, we can easily compute the singular part of the amplitudes 
from the known results of ref.\cite{BERN1}
by using the procedure given in ref.\cite{SIGN,BILL}. 
To evaluate the singular part, we have to keep in mind that 
the regularization scheme does not affect 
the universal structure of the $1/\epsilon^2$ pole parts  
but the $1/\epsilon$ pole parts are scheme dependent. 
We estimated the singular parts in the FDH scheme 
and our results are consistent with the results 
of ref.\cite{SIGN,BILL}.

We also performed a consistency check of the 
scheme dependence. 
It is well known that, the FDH scheme at one-loop level 
is equivalent to the dimensional reduction(DR) scheme\cite{SIEG}. 
The conversion relation between the DR scheme 
and 't Hooft-Veltman scheme for the two-quark (n-2)gluon amplitudes 
is discussed in ref.\cite{BERN1,KUN2}. 
From their argument, 
the conversion from the DR scheme to the 't Hooft-Veltman 
scheme at one-loop level 
is obtained by shifting the amplitudes as 
$A_n \rightarrow A_n+\delta_n$ with   
$\delta_n=-c_\Gamma (1-{1\over N_c^2})A_n^{tree}$. 
Now, we can consider the color-less limits 
($C_f\rightarrow 1$ and $C_A\rightarrow 0$) 
of our results and the results in ref.\cite{BERN1,BILL}. 
In this limits, both results give the same 
quantity of shifting parameter $\delta_5$.

%\newpage
%\input chap4.tex
\section{Conclusion}

In this paper, 
we presented one-loop five-parton  amplitudes involving 
two massless quarks, two or three photons external legs. 
The one-loop five parton amplitudes with external photons 
are required to evaluate the QCD background for 
the Higgs production in association with a jet. 
These amplitudes have been discussed\cite{SIGN,BILL} 
in terms of the known QCD amplitude($q\bar{q}ggg $)\cite{BERN1}. 
They give the systematic procedure to replace external 
gluons into photons. However, their results are still 
in rather large expression.
Here, we computed the one-loop helicity amplitudes 
for the process $q\bar{q}g\gamma\gamma$ directly 
and obtained more compact expression successfully. 
The amplitudes which we presented here, together 
with the six parton tree-level amplitudes 
($q\bar{q}\gamma\gamma gg$,
$q\bar{q}Q\bar{Q}\gamma\gamma$) 
\cite{BILL}, 
we can estimate the NLO QCD background for 
the associated Higgs production with a jet. 

\vspace{1cm}
{\bf Acknowledgments}

I would like to thank Prof. Bern and Dr. Kilgore for 
useful discussions and comments, and 
Prof. Tr\'ocs\'anyi for pointing out an error
in an earlier version of eq.(\ref{FPA}) 
and typos. 
I also thank Prof. Okada and Prof. Yazaki. 

%\newpage
%\input ref.tex

\end{document}